\documentclass[aps,reprint,twocolumn,showpacs,superscriptaddress]{revtex4-1}
\usepackage{graphicx}
\usepackage{xcolor}
\usepackage{epstopdf}
\usepackage{dcolumn}
\usepackage{bm}
\usepackage{color}
\usepackage{subfigure}

\begin{document}


\title{Line nodes and surface Majorana flat bands in static and kicked $p$-wave superconducting Harper model}

\author{Huai-Qiang Wang}
\affiliation{National Laboratory of Solid State Microstructures
and School of Physics, Nanjing University, Nanjing 210093, China}

\author{M.N. Chen}
\affiliation{National Laboratory of Solid State Microstructures
and School of Physics, Nanjing University, Nanjing 210093, China}

\author{Raditya Weda Bomantara}
\affiliation{Department of Physics, National University of Singapore, Singapore 117543}

\author{Jiangbin Gong}
\email{phygj@nus.edu.sg}
\affiliation{Department of Physics, National University of Singapore, Singapore 117543}

\author{D.Y. Xing}
\email{dyxing@nju.edu.cn}
\affiliation{National Laboratory of Solid State Microstructures and School of Physics, Nanjing
University, Nanjing 210093, China} \affiliation{Collaborative
Innovation Center of Advanced Microstructures, Nanjing University,
Nanjing 210093, China}





\begin{abstract}
We investigate the effect of introducing nearest-neighbor $p$-wave superconducting pairing to both the static and kicked extended Harper model with two periodic phase parameters acting as artificial dimensions to simulate three-dimensional systems. It is found that in both the static model and the kicked model, by varying the $p$-wave pairing order parameter, the system can switch between a fully gapped phase and a gapless phase with point nodes or line nodes. The topological property of both the static and kicked model is revealed by calculating corresponding topological invariants defined in the one-dimensional lattice dimension. Under open boundary conditions along the physical dimension, Majorana flat bands at energy zero (quasienergy zero and $\pi$) emerge in the static (kicked) model at the two-dimensional surface Brillouin zone. For certain values of pairing order parameter, (Floquet) Su-Schrieffer-Heeger-like edge modes appear in the form of arcs connecting different (Floquet) Majorana flat bands. Finally, we find that in the kicked model, it is possible to generate two controllable Floquet Majorana modes, one at quasienergy zero and the other at quasienergy $\pi$, at the same parameter values.
\end{abstract}

\pacs{03.65.Vf, 73.43.-f, 74.20.Rp,  71.10.Pm}

\maketitle

\section{Introduction}
The last decade has seen tremendous advances in our understandings of topological phases, ranging from gapped phases, such as topological insulators and topological superconductors~\cite{RevModPhys.82.3045,RevModPhys.83.1057}, to gapless topological semimetals, such as Weyl semimetals~\cite{PhysRevB.83.205101,PhysRevLett.107.127205,Xu2015TOPOLOGICAL,PhysRevX.5.011029} and nodal-line semimetals~\cite{PhysRevB.84.235126,PhysRevB.90.115111,PhysRevB.92.081201,PhysRevB.93.121113,
PhysRevLett.115.026403,PhysRevLett.115.036806,PhysRevLett.115.036807}. Both fully gapped and gapless topological materials can be classified in terms of nonspatial symmetries, such as time reversal and particle hole, as well as spatial symmetries, such as reflection and rotation~\cite{PhysRevB.78.195125, RevModPhys.88.035005}. Each topological class can be characterized by a topological invariant calculated from its bulk spectrum, which cannot change without the closing-reopening process of the bulk gap. Through the bulk-edge correspondence, when open boundary conditions (OBCs) are taken, gapless edge (surface) states will emerge, such as chiral (helical) edge states in quantum (spin) Hall insulators, Majorana zero modes in topological superconductors~\cite{RevModPhys.82.3045, RevModPhys.83.1057}, Fermi arcs in Weyl semimetal~\cite{PhysRevB.83.205101,Xu2015TOPOLOGICAL}, and surface flat bands in nodal-line semimetals~\cite{PhysRevB.84.235126,PhysRevB.90.115111,PhysRevB.92.081201,PhysRevB.93.121113,
PhysRevLett.115.026403,PhysRevLett.115.036806,PhysRevLett.115.036807}.

Recently, topological phases in periodically driven quantum systems have also attracted considerable theoretical and experimental interest due to their high controllability~\cite{Lindner2010Floquet, PhysRevB.79.081406, PhysRevB.82.235114,
 PhysRevB.84.115133, PhysRevB.87.201109, PhysRevB.87.235131, PhysRevB.88.155133, PhysRevB.91.085420,PhysRevB.90.195419, PhysRevE.93.022209, PhysRevLett.105.017401, PhysRevLett.106.220402, PhysRevLett.107.216601, PhysRevLett.108.056602, PhysRevLett.110.026603, PhysRevLett.110.200403, PhysRevLett.111.047002, PhysRevLett.112.026805, PhysRevX.3.031005, 0295-5075-105-1-17004, PhysRevLett.116.176401, Manisha2016Floquet, PhysRevB.93.075405, PhysRevB.94.205429, PhysRevB.93.115429, 1367-2630-17-12-125014, PhysRevB.90.125143, Roy2016Floquet}.  An otherwise topologically trivial system can be made to be topologically nontrivial by means of a driving field, with one seminal example termed as ``Floquet topological insulator'' (FTI)~\cite{Lindner2010Floquet} in connection with the Floquet theory \cite{PhysRev.138.B979, PhysRevA.7.2203}. Unlike static systems, in a periodically driven (hence Floquet) system,  the quasienergy is defined only up to a Brillouin zone (BZ).  As such, two kinds of bulk Floquet gaps exist, i.e., one around quasienergy zero and the other around $\pi$, leading to two types of edge modes, namely, the zero mode and $\pi$ mode. Floquet superconductors were first proposed in Ref.~\cite{PhysRevLett.106.220402}, where a $Z_{2}\times Z_{2}$ invariant was introduced to describe the zero-mode as well as $\pi$-mode Floquet Majorana fermions (FMFs). In addition to gapped Floquet phases, there are also proposals to realize Floquet topological semimetals~\cite{0295-5075-105-1-17004, PhysRevE.93.022209, PhysRevLett.117.087402, PhysRevB.93.144114}. For example, Weyl semimetals can be induced from three-dimensional (3D) topological insulators~\cite{0295-5075-105-1-17004} or nodal-line semimetals~\cite{PhysRevLett.117.087402} by the application of off-resonant circularly polarized light. There are many interesting effects peculiar to periodically driven systems such as the anomalous edge states~\cite{PhysRevX.3.031005} and counterpropagating chiral edge modes~\cite{PhysRevLett.112.026805}, which are still under investigation.

 Among the studies of Floquet topological phases, the kicked Harper model (KHM) provides an opportunity to explore topological phenomena in higher-dimensional systems from the perspective of one-dimensional (1D) systems by introducing periodic system parameters as artificial dimensions~\cite{PhysRevB.90.195419, PhysRevE.93.022209, PhysRevB.94.075443}. For example, in Ref.~\cite{PhysRevE.93.022209}, modulations were introduced to both the lattice hopping term and the kicking potential in the KHM, which were represented by two phase shifts playing the role of two periodic system parameters. By tuning the hopping strength and kicking strength in the system, 3D Floquet Weyl semimetals and nodal-line semimetals could be easily realized. Motivated by these results and the property of FMFs~\cite{PhysRevLett.106.220402}, in this paper, nearest-neighbor $p$-wave superconducting pairing in the lattice dimension is introduced into the above extended KHM to explore 3D Floquet topological superconducting phases. In order to capture unique features of Floquet topological superconducting phases, we also study the static version of the model to make a comparison.

The main results of this paper are as follows. First, it is found that in both the static model and the kicked model, the introduction of $p$-wave superconductivity may give rise to point nodes, line nodes, or a full gap in the bulk (quasi) energy spectrum. While the static model can only exhibit point or line nodes at energy zero, the kicked model can exhibit point or line nodes at both quasienergy zero and $\pi$. As the pairing order parameter is increased, the line nodes expand in the static case, which eventually merge with one another and disappear. In the kicked case, however, as the pairing order parameter is increased beyond a certain value, additional band touching points emerge. Second, by treating the two phase shifts as parameters, topological $Z$ and $Z_{2}$ indices can be calculated in the 1D lattice dimension for the static model, which are extended to $Z\times Z$ and $Z_{2}\times Z_{2}$ indices for the kicked model due to the extra $\pi$ mode. When OBCs are taken along the lattice dimension, zero energy Majorana flat bands (two flavors of Floquet Majorana flat bands at quasienergy zero and $\pi$) emerge at the 2D surface BZ spanned by the two periodic parameters in the static (kicked) model, which can be characterized by the $Z_2$ ($Z_2\times Z_2$) index and bounded by the line nodes. Due to the controllability of the line nodes, it is possible to turn the whole parameter region into a topologically nontrivial (quasi)-band structure with a full bulk (quasi)-energy gap, leading to the emergence of flat bands over the whole parameter BZ under OBCs. These flat bands contain a large density of states for (Floquet) Majorana fermions (MFs) and, due to their controllability, may provide a versatile platform to study various properties of MFs such as interaction effects~\cite{1367-2630-15-8-085002, PhysRevB.89.174510} and transport phenomena. Third, at certain values of pairing order parameter, (Floquet) Su-Schrieffer-Heeger (SSH)-like edge modes, i.e., flat lines connecting different nodal loops, emerge at the surface and coexist with the (Floquet) Majorana modes. This implies that both (Floquet) SSH-like edge modes and (Floquet) Majorana edge modes can be realized in the same system.  Finally, a unique feature of the kicked model, i.e., the possibility to generate two controllable Floquet Majorana modes, one at quasienergy zero and the other at quasienergy $\pi$, at the same parameter values, is again observed.

This paper is organized as follows. In Sec. II, we first present the Hamiltonian of the static superconducting off-diagonal Harper model (SODHM) made up by adding the nearest-neighbor $p$-wave pairing term to the extended off-diagonal Harper model~\cite{PhysRevLett.110.180403}. The condition for the emergence of line nodes is derived in Sec. II A. The $Z$ and $Z_{2}$ topological invariant is introduced and calculated in Sec. II B, and the surface flat bands under OBCs are shown numerically in Sec. II C. We then continue to investigate the kicked superconducting off-diagonal Harper model (KSODHM) in Sec. III. The Floquet spectrum is derived and compared with its static counterpart in Sec. III A. The $Z_{2}\times Z_{2}$ index is introduced and calculated in Sec. III B. The Floquet surface flat bands under OBCs are shown and a unique feature of the kicked model is discussed in Sec. III C.  Section IV concludes this paper.

\section{Static superconducting off-diagonal Harper model (SODHM)}
\subsection{Model description}
We first investigate the static Harper model (HM) with modulated off-diagonal hopping terms, modulated on-site potentials, and nearest-neighbor $p$-wave pairing terms. The Hamiltonian is given as
\begin{eqnarray}
 \label{eqn.1}
   H(t)=&&\sum_{n=1}^{N-1}\Big\{[J+\lambda\cos(2\pi\alpha_{1}n+\phi_{y})]\hat{c}_{n}^{\dagger}\hat{c}_{n+1}\nonumber\\
    &&+\Delta \hat{c}_{n}^{\dagger}\hat{c}_{n+1}^{\dagger}+\mathrm{H.c.}\Big\}
    +\sum_{n=1}^{N}V\cos(2\pi\alpha_{2}n+\phi_{z})\hat{c}_{n}^{\dagger}\hat{c}_{n},\nonumber\\
\end{eqnarray}
where $N$ counts the total number of lattice sites, $\hat{c}_{n}^{\dagger}(\hat{c}_{n})$ denotes the creation (annihilation) operator, $J$ and $\lambda$ are nearest-neighbor hopping strength, $\Delta$ denotes the nearest-neighbor $p$-wave pairing order parameter ($\Delta>0$ is assumed hereafter), $V$ is the on-site potential strength, $\alpha_{1}$ and $\alpha_{2}$ are two adjustable parameters controlling the nearest-neighbor hopping and on-site potential terms, respectively, with $\phi_{y}$ and $\phi_{z}$ representing the hopping and kicking phase shifts. As in Ref.~\cite{PhysRevE.93.022209}, for simplicity, we take $\alpha_{1}=\alpha_{2}=\frac{1}{2}$ in Eq.~(1) and get
\begin{eqnarray}
 \label{eqn.2}
   H=&&\sum_{n=1}^{N-1}\{[J+(-1)^n\lambda\cos\phi_{y}]\hat{c}_{n}^{\dagger}\hat{c}_{n+1}+\Delta \hat{c}_{n}^{\dagger}\hat{c}_{n+1}^{\dagger}+\mathrm{H.c.}\}\nonumber\\
   &&+\sum_{n=1}^{N}(-1)^nV\cos\phi_{z}\hat{c}_{n}^{\dagger}\hat{c}_{n}.
\end{eqnarray}
Since the phases $\phi_{y}$ and $\phi_{z}$ are also periodic, they can be treated as quasimomenta perpendicular to the physical lattice direction. The system is invariant after translation by two lattice sites under periodic boundary conditions in the lattice direction, so Fourier transform can be performed with two sublattices in a unit cell,
\begin{eqnarray}
\label{eqn.3}
\hat{c}_{s,n}=\frac{1}{\sqrt{N/2}}\sum_{k\in[0,\pi]}\hat{c}_{s,k}e^{ik(2n+s-2)}
\end{eqnarray}
where $s=1,2$ for odd and even lattice sites, respectively. Note that because of the enlarged unit cell, the BZ has now shrunk into half of the initial one. Under this transformation, the quasimomentum space single-particle Hamiltonian can be written as
\begin{eqnarray}
 \label{eqn.4}
 h(k,\phi_{y},\phi_{z})=&&J\cos (k)\tau_{z}\sigma_{x}+\lambda\cos(\phi_{y})\sin (k)\tau_{z}\sigma_{y}
 \nonumber\\
 &&+\Delta\sin (k)\tau_{y}\sigma_{x}+\frac{V}{2}\cos(\phi_{z})\tau_{z}\sigma_{z},
\end{eqnarray}
where the Pauli matrices $\sigma_i$ and $\tau_i$, with $i=0,x,y,z$, act in the sublattice space and particle-hole space, respectively, and $[c_{1,k},c_{2,k},c_{1,-k}^{\dagger},c_{2,-k}^{\dagger}]^{T}$ is chosen as the base vector. At $k=\frac{\pi}{2}$, the energy spectrum is given by $\epsilon=\pm\Delta\pm\sqrt{\lambda^{2}\cos^{2}\phi_{y}+\frac{V^{2}\cos^{2}\phi_{z}}{4}}$. The gap closing condition corresponds to $\epsilon=0$, which is satisfied when
\begin{equation}
\label{eqn.5}
\Delta=\sqrt{\lambda^{2}\cos^{2}\phi_{y}+\frac{V^{2}\cos^{2}\phi_{z}}{4}}.
\end{equation}
If $\Delta=0$, the middle two bands touch at isolated points $(k,\phi_{y},\phi_{z})=(\frac{\pi}{2},\pm\frac{\pi}{2},\pm\frac{\pi}{2})$, while if $\Delta\neq0$, they touch along nodal loops or lines determined by Eq.~(5), as explicitly illustrated in Fig.~1(a) with $J=1,\lambda=0.5,V=2$ and different values of $\Delta$. If $\Delta=\sqrt{\lambda^{2}+\frac{V^2}{4}}$, they again touch at isolated points
$(k,\phi_{y},\phi_{z})=\Big(\frac{\pi}{2},0(\pi),0(\pi)\Big)$. Finally, if $\Delta>\sqrt{\lambda^{2}+\frac{V^2}{4}}$, there will be no band touching points, resulting in a total gap between particle bands and hole bands.
\begin{figure}
  \centering
\includegraphics[scale=0.5]{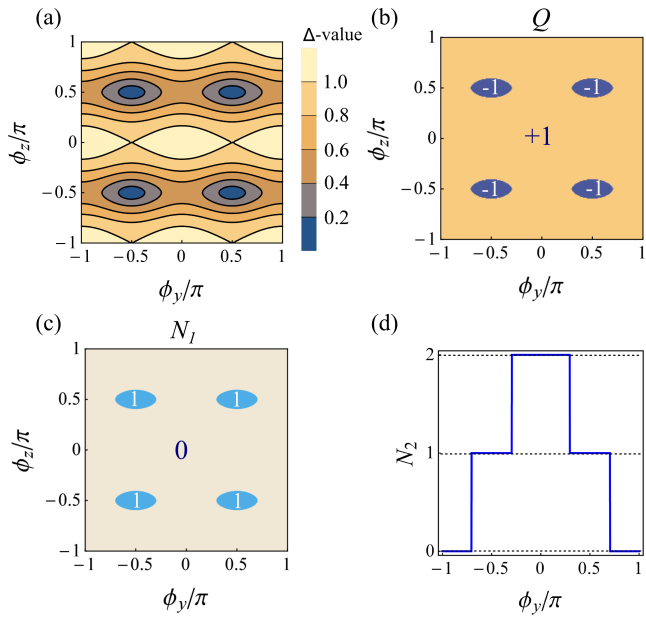}
  \caption{(Color online) (a) Nodal lines as a function of $\phi_{y}$ and $\phi_{z}$ in the $k=\frac{\pi}{2}$ plane for $J=1,\lambda=0.5,V=2$ with different values of $\Delta$.  (b) The $Z_{2}$ index, (c) the chiral index $N_{1}$, and (d) $N_{2}$ for $J=1,\lambda=0.5,V=2$, and $\Delta=0.3$.}
  \label{Fig1}
\end{figure}

\subsection{$Z$ and $Z_{2}$ topological invariant}
Due to the nearest-neighbor $p$-wave superconductor pairing, the static Hamiltonian in Eq.~(4) possesses particle-hole symmetry (PHS), $Ph(k,\phi_{y},\phi_{z})P^{-1}=-h(-k,-\phi_{y},-\phi_{z})$, with the particle-hole operator $P=\tau_{x}K$, where $K$ denotes the complex conjugate. The Hamiltonian also satisfies the time-reversal symmetry (TRS), $Th(k,\phi_{y},\phi_{z})T^{-1}=h(-k,-\phi_{y},-\phi_{z})$, with the time-reversal operator defined as $T=K$ for a spinless system. Consequently, there exists a chiral symmetry (CS) in the system, $C_{1}h(k,\phi_{y},\phi_{z})C^{-1}_{1}=-h(k,\phi_{y},\phi_{z})$, with the chiral operator defined as the product of the time-reversal operator and particle-hole operator $C_{1}=TP=\tau_{x}$. When $\phi_{z}=\pm\frac{\pi}{2}$, the system has an additional SSH-like sublattice(chiral) symmetry, $C_{2}h(k,\phi_{y},\phi_{z})C^{-1}_{2}=-h(k,\phi_{y},\phi_{z})$, with $C_{2}=\sigma_{z}$ \cite{PhysRevLett.42.1698, PhysRevB.90.014505, Zeng2016}. According to the Altland-Zirnbauer classification for noninteracting systems~\cite{PhysRevB.78.195125}, this Hamiltonian belongs to the BDI class. However, in three dimensions, no strong topological invariants exist to characterize the BDI class. Nevertheless, when $\phi_{y}$ and $\phi_{z}$ are treated as sole parameters, the system will be reduced to 1D chains of the BDI class, which can now be characterized by two $Z$ chiral indices corresponding to the above two CS, respectively. For general $\phi_{y}$ and $\phi_{z}$ with CS $C_{1}$, the chiral index $N_{1}$ is given by (see Appendix A)~\cite{PhysRevB.90.014505,Zeng2016}
\begin{eqnarray}
\label{eqn.6}
N_{1}=\left\{ \begin{array}{ll}
1& \textrm{for $\Delta>\sqrt{\lambda^{2}\cos^{2}\phi_{y}+\frac{V^{2}\cos^{2}\phi_{z}}{4}}$},\\
0 & \textrm{for $\Delta<\sqrt{\lambda^{2}\cos^{2}\phi_{y}+\frac{V^{2}\cos^{2}\phi_{z}}{4}}$}.
\end{array}\right.
\end{eqnarray}
For the sublattice symmetric case of $\phi_{z}=\pm\frac{\pi}{2}$, another chiral index $N_{2}$ can be obtained as (see Appendix A)
\begin{eqnarray}
\label{eqn.7}
N_{2}&=&\Theta(\lambda\cos\phi_{y}-\Delta)+\Theta(\lambda\cos\phi_{y}+\Delta)\nonumber\\
&=&\left\{ \begin{array}{ll}
2& \textrm{for $\lambda\cos\phi_{y}>\Delta$},\\
1& \textrm{for $-\Delta<\lambda\cos\phi_{y}<\Delta$},\\
0 & \textrm{for $\lambda\cos\phi_{y}<-\Delta$}.
\end{array}\right.
\end{eqnarray}
A nonzero $N_{1}$ or $N_{2}$ indicates a topological nontrivial phase. Apart from the two $Z$ indices, there also exists a weak particle-hole $Z_{2}$ index~\cite{Kitaev, PhysRevLett.109.150408, Wang2016Flux}, which is obtained as (see Appendix B)
\begin{equation}
\label{eqn.8}
  Q=\mathrm{sgn}\bigg\{\bigg(\lambda^{2}\cos^{2}\phi_{y}+\frac{V^{2}\cos^{2}\phi_{z}}{4}\bigg)-\Delta^{2}\bigg\}.
\end{equation}
Here, $Q=-1(1)$ stands for $Z_{2}$ nontrivial (trivial) phase, and $Q$ is related to $N_{1}$ through $Q=(-1)^{N_{1}}$, which means the parity of $N_{1}$. In the current model without longer-ranged-neighbor couplings , $N_{1}$ can-not be larger than $1$, so the nontrivial region with $N_{1}=1$ should coincide with that with a nontrivial $Q=-1$. By comparing Eqs.~(6)--(8) with Eq.~(5), it is easy to see that these nodal lines correspond to boundaries between topological nontrivial and trivial phases, which is quite reasonable since the change of topological invariants usually requires the closing-reopening process of the bulk gap. We choose $J=1,\lambda=0.5,V=2$, and $\Delta=0.3$, and numerically show $Q$, $N_{1}$, and $N_{2}$ in Figs.~1(b)--1(d), respectively, where blue regions of $N_{1}=1$ in Fig.~1(c) [or $Q=-1$ in Fig.~1(b)] and $N_{2}\neq0$ sections in Fig.~1(d) represent nontrivial phases. When $\Delta>\sqrt{\lambda^{2}+\frac{V^{2}}{4}}$, no nodal points or lines exist and $N_{1}$ and $N_{2}$ always equal $1$, indicating a nontrivial phase regardless of the values of $\phi_{y}$ and $\phi_{z}$. Because of the bulk-edge correspondence, as is elucidated in Sec. II C, this implies the existence of one Majorana zero mode throughout the whole parameter BZ.

\subsection{Surface Majorana flat bands and Dirac arcs under OBCs}
It is well known that, through bulk-edge correspondence, topological phases can also be characterized by their localized edge states when OBCs are taken, such as chiral (helical) edge states in quantum (spin) Hall insulators, Majorana zero modes in topological superconductors~\cite{RevModPhys.82.3045,RevModPhys.83.1057}, and Fermi arcs in Weyl semimetals~\cite{PhysRevB.83.205101}. Similarly, for each of the above 1D chains with a nontrivial $N_{1}$ or $Z_{2}$ index, Majorana zero modes will emerge at the chain ends when OBCs are taken along the lattice direction. As a result, surface Majorana flat bands will emerge in the 2D surface BZ. This is quite similar to the drumhead surface states in nodal-line semimetals which are protected by the $Z_{2}$ parity of Berry phase~\cite{PhysRevLett.115.036807,PhysRevLett.115.036806}, and surface flat bands in other systems~\cite{Heikkil2011}.
\begin{figure}
  \centering
  \includegraphics[scale=0.65]{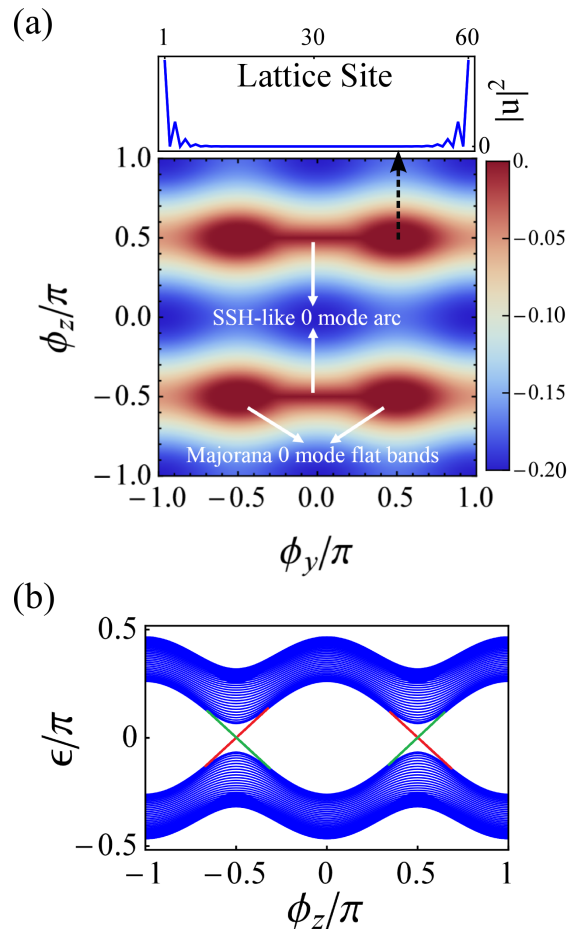}
  \caption{(Color online) For fixed $J=1,\lambda=0.5,V=2$, and $\Delta=0.3$ under OBCs in the lattice dimension with 60 lattice sites , (a) the lower one of the middle two energy bands around zero energy as a function of $\phi_{y}$ and $\phi_{z}$, where Majorana zero-mode flat bands as well as Dirac zero-mode arcs connecting them can be seen clearly. Inset: The modular square of the wavefunction's particle component against the lattice sites for the state $(\phi_{y},\phi_{z})=(\frac{\pi}{2},\frac{\pi}{2})$ inside the flat bands. (b) The energy spectrum as a function of $\phi_{z}$ for fixed $\phi_{y}=0$, where counterpropagating chiral edge states appear around $\phi_{z}=\pm\frac{\pi}{2}$.}
  \label{Fig2}
\end{figure}

Figure 2 shows the energy spectrum of the static model under OBCs with $N=60$ lattice sites and $J=1,\lambda=0.5, V=2$, $\Delta=0.3$. The lower one of the middle two bands around $\epsilon=0$ is shown in Fig.~2(a), where Majorana zero-energy flat bands are clearly seen inside the nodal loops determined by the projections of the bulk bands onto the surface BZ. To see the localized nature of these flat bands, we write an energy eigenstate in the real space as $\hat{\psi}=\sum^N_{i=1}\big(u_{i}\hat{c}_{i}+v_{i}\hat{c}^{\dag}_i\big)$, where $u_{i}(v_{i})$ stands for particle (hole) component. As shown in the inset of Fig.~2(a), we choose one state $(\phi_{y},\phi_{z})=(\frac{\pi}{2},\frac{\pi}{2})$ inside the flat bands and plot the modular square of its particle component against the lattice sites $|u_{i}|^2$, which decays fast into the bulk and is truly localized at the surface.

Apart from the drumhead states, when $\Delta<|\lambda|$, there exist two arcs at $\phi_{z}=\pm\frac{\pi}{2}$ connecting two flat bands, as can be clearly seen in Fig.~2(a). The emergence of the arcs can be explained by the additional chiral (sublattice) symmetry $C_{2}$ when $\phi_{z}=\pm\frac{\pi}{2}$ in Eq.~(4) and share the same origin as those in the SSH model~\cite{PhysRevB.90.014505, PhysRevE.93.022209, PhysRevLett.42.1698,Zeng2016}. For these values of $\phi_{y}$ and $\phi_{z}$, $N_{2}=2$, indicating that each of these modes is a Dirac zero mode formed by two Majorana zero modes. These arcs can also be explained as follows. By scanning $\phi_z$ from $-\pi$ to $\pi$ at a fixed value of $\phi_y$ between $-\frac{\pi}{2}$ and $\frac{\pi}{2}$, counterpropagating chiral edge states, i.e., one around $\phi_{z}=\frac{\pi}{2}$, and the other around $\phi_{z}=-\frac{\pi}{2}$, traverse the bulk gap with opposite velocity, as shown in Fig.~2(b). These counterpropagating chiral edge states originate from the band structure of the system at a fixed $|\phi_y|<\frac{\pi}{2}$. When $\Delta=0$, the band structure for a 2D $k-\phi_{z}$ slice at a constant $|\phi_y|<\frac{\pi}{2}$ plane corresponds to a topologically nontrivial insulator formed by two mirror copies of Chern insulators due to the presence of TRS~\cite{PhysRevB.91.125438}, while a 2D $k-\phi_{z}$ slice outside that range or a 2D $k-\phi_{y}$ slice corresponds to trivial insulators with no chiral edge states. When a small $\Delta$ term is introduced, as long as the bulk gap on this plane does not close and can be adiabatically connected to the $\Delta=0$ case, these counterpropagating chiral edge states should remain, since they are topologically protected. The crossings of these chiral edge states at zero energy form the arcs connecting different nodal loops. This model therefore offers an opportunity to realize both the SSH-like Dirac zero mode and the Kitaev-like Majorana zero mode, which have different topological origins, within the same system.
\section{kicked superconducting off-diagonal Harper model (KSODHM)}
\subsection{Model description}
We now consider the kicked version of the SODHM by replacing the static on-site potential in Eq.~(2) with a periodically kicked one, and rewrite Eq.~(2) as
\begin{eqnarray}
 \label{eqn.9}
   H=&&\sum_{n=1}^{N-1}\{[J+(-1)^n\lambda\cos\phi_{y}]\hat{c}_{n}^{\dagger}\hat{c}_{n+1}+\Delta \hat{c}_{n}^{\dagger}\hat{c}_{n+1}^{\dagger}+\mathrm{H.c.}\}\nonumber\\
   &&+\sum_{n=1}^{N}\sum_{m}(-1)^nV\delta(t-mT)\cos\phi_{z}\hat{c}_{n}^{\dagger}\hat{c}_{n},
\end{eqnarray}
where $T$ represents the kicking period. Due to the time-periodic nature of the above Hamiltonian, we can apply Floquet theory and define the Floquet operator for one full period as~\cite{Lindner2010Floquet,PhysRevE.93.022209}
\begin{eqnarray}
 \label{eqn.10}
 U(T)=&&e^{-i\sum_{n=1}^{N-1}\{[J+(-1)^n\lambda\cos\phi_{y}]\hat{c}_{n}^{\dagger}\hat{c}_{n+1}+\Delta \hat{c}_{n}^{\dagger}\hat{c}_{n+1}^{\dagger}+\mathrm{H.c.}\}}\nonumber\\
 &&\times e^{-i\sum_{n=1}^{N}(-1)^nV\cos\phi_{z}\hat{c}_{n}^{\dagger}\hat{c}_{n}},
\end{eqnarray}
where $\hbar=T=1$ have been taken for simplicity. Through the same Fourier transform as introduced in Eq.~(3), the quasimomentum space Floquet operator $U(k,\phi_{y},\phi_{z})$ is given by
\begin{eqnarray}
 \label{eqn.11}
 U(k,\phi_{y},\phi_{z})=e^{-ih_{0}}e^{-ih_{1}},
\end{eqnarray}
with
\begin{eqnarray}
\label{eqn.12}
 h_{0}&=&J\cos (k)\tau_{z}\sigma_{x}+\lambda\cos(\phi_{y})\sin (k)\tau_{z}\sigma_{y}
 +\Delta\sin (k)\tau_{y}\sigma_{x}\nonumber\\
 h_{1}&=&\frac{V}{2}\cos(\phi_{z})\tau_{z}\sigma_{z}.
\end{eqnarray}
The quasienergy spectrum can be obtained from the exponents of the eigenvalues of the Floquet operator~\cite{PhysRev.138.B979,PhysRevA.7.2203}. Without superconductivity, i.e., if $\Delta=0$, this model will be reduced to the one in Ref.~\cite{PhysRevE.93.022209}.

In order to further simplify Eq.~(11), we shall focus on the plane $k=\frac{\pi}{2}$ in the 3D BZ. By performing a cyclic permutation $\tau_{x}\rightarrow\tau_{z}\rightarrow\tau_{y}\rightarrow\tau_{x}$, the argument in both exponential operators is block diagonal and can be written in the form
\begin{eqnarray}
\label{eqn.13}
 A=\left(
  \begin{array}{cc}
    0 & a \\
    a^{\dagger} & 0 \\
  \end{array}
\right),
\end{eqnarray}
where $a$ is a $2\times2$ matrix. In the new representation of $\tau's$, the argument in the first exponent of Eq.~(11) is given by Eq.~(13) with $a=-i\lambda\cos (\phi_{y})\sigma_{y}+\Delta\sigma_{x}$. It follows that $A^{2n}=[\Delta-\lambda\cos(\phi_{y})\tau_{z}\sigma_{z}]^{2n}$ and $A^{2n-1}=[\Delta-\lambda\cos(\phi_{y})\tau_{z}\sigma_{z}]^{2n-1}\tau_{x}\sigma_{x}$. Therefore, by using the property of exponential expansion,
\begin{eqnarray}
\label{eqn.14}
e^{-i[\lambda\cos(\phi_{y})\tau_{y}\sigma_{y}+\Delta\tau_{x}\sigma_{x}]}=&&\cos\big(\Delta-\lambda\cos(\phi_{y})\tau_{z}\sigma_{z}\big)\nonumber\\
&&-i\sin\big(\Delta-\lambda\cos(\phi_{y})\tau_{z}\sigma_{z}\big)\tau_{x}\sigma_{x}.\nonumber\\
\end{eqnarray}
Similarly, for the second exponent of Eq.~(11), $a=-i\frac{V}{2}\cos(\phi_{z})\sigma_{z}$, so that $A^{2n}=(\frac{V}{2}\cos\phi_{z})^{2n}I_{4}$ and $A^{2n-1}=(\frac{V}{2}\cos\phi_{z})^{2n-1}\tau_{y}\sigma_{z}$. Therefore,
\begin{eqnarray}
\label{eqn.15}
e^{-i\frac{V\cos\phi_{z}}{2}\tau_{y}\sigma_{z}}=\cos\bigg(\frac{V\cos\phi_{z}}{2}\bigg)-i\sin\bigg(\frac{V\cos\phi_{z}}{2}\bigg)\tau_{y}\sigma_{z}.\nonumber\\
\end{eqnarray}

Equation (5), together with Eqs.~(14) and (15), lead to the following approximate condition for the band touching points (on the $k=\frac{\pi}{2}$ plane) to occur:
\begin{equation}
\label{eqn.16}
\Delta (\mathrm{mod} \;\pi)=\sqrt{\lambda^{2}\cos^{2}\phi_{y}+\left[\frac{V\cos\phi_{z}}{2}(\mathrm{mod}\;\pi)\right]^{2}},
\end{equation}
which agrees quite well with numerical results for small parameters of $\lambda, V, \Delta <1$. For larger values of these parameters, there are in general some significant discrepancies between Eq.~(16) and our numerics. However, Eq.~(16) becomes exact and can be verified analytically in the special cases of $\lambda\cos\phi_{y}=0$ or $\frac{V}{2}\cos\phi_{z}=l\pi$ ($l$ is an integer). If $\frac{V}{2}\cos\phi_{z}=l\pi$, band touchings occur when $\Delta=\lambda\cos\phi_{y}+m\pi$, where $m$ is an integer, in which case the Floquet operator becomes
\begin{eqnarray}
\label{eqn.17}
U=&&(-1)^{l+m}\times\nonumber\\
&&\Big\{\cos[\Delta(1-\tau_{z}\sigma_{z})]-i\sin[\Delta(1-\tau_{z}\sigma_{z})]\tau_{x}\sigma_{x}\Big\}.\nonumber\\
\end{eqnarray}
It is easily verified that Eq.~(16) has two degenerate eigenvalues $(-1)^{l+m}$, which means that two quasienergies touch at zero $(\pi)$ for even (odd) $l+m$. If $\lambda\cos\phi_{y}=0$, then $\frac{V}{2}\cos\phi_{z}=\Delta+l\pi$ is required to achieve band touching, in which case,
\begin{eqnarray}
\label{eqn.18}
U=&&(-1)^{l}\times \nonumber\\
&&\Big[\cos^{2}\Delta-\sin^{2}(\Delta)\tau_{z}\sigma_{y}
-\frac{i\sin(2\Delta)}{2}(\tau_{y}\sigma_{z}+\tau_{x}\sigma_{x})\Big],\nonumber\\
\end{eqnarray}
which possesses two degenerate eigenvalues $(-1)^{l}$, giving rise to band touchings at quasienergy zero $(\pi)$ for even (odd) $l$.

Equations (17) and (18) provide an analytical argument that, in the kicked model, band touching points can occur not only at quasienergy zero, but also at quasienergy $\pi$. To further compare the bulk spectrum of the static and kicked model, we consider the same parameters as those in the static case $J=1,\lambda=0.5$, and $V=2$. If $\Delta=0$, band touchings occur only at quasienergy zero at $(k,\phi_{y},\phi_{z})=\left(\frac{\pi}{2},\pm\frac{\pi}{2}, \pm\frac{\pi}{2}\right)$ \cite{PhysRevE.93.022209}, as shown in Fig.~3(a). After turning on a nonzero $\Delta$, as can be seen in Fig.~3(b) with $\Delta=0.3$, the upper and lower branches begin to split apart into two particle bands and two hole bands, respectively, and they touch along closed nodal lines instead of isolated points on the $k=\frac{\pi}{2}$ plane. $\Delta=0.5$ and $\Delta=1$ correspond to the two special cases where different nodal lines start to touch and merge with each other, as shown in Fig.~3(c). When $\Delta\approx1.1$, all the nodal lines have merged with each other and have vanished to form a total gap. So far, everything is similar to the static model; however, if we continue to increase $\Delta$ to a value of $\Delta \approx 2.07$, new band touching points, now at quasienergy $\pi$, start to emerge, as shown in Fig.~3(d). These additional band touching points lead to a different topology and a different feature of the kicked model, which will be discussed in Secs. III B and III C.

\begin{figure}
  \centering
  \includegraphics[scale=0.6]{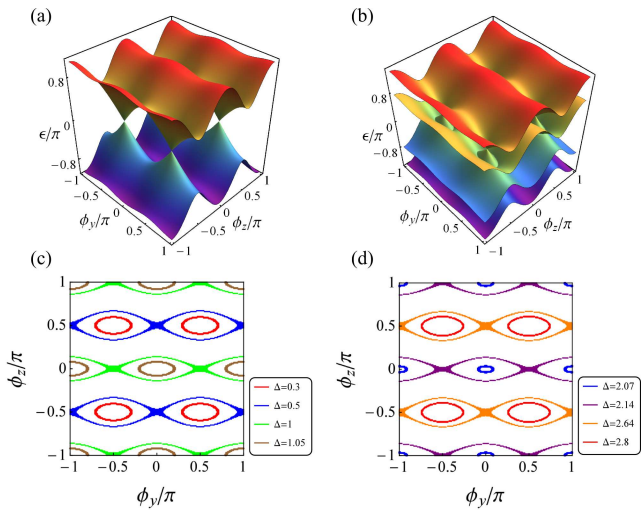}
  \caption{(Color online) The quasienergy spectrum as a function of $\phi_{y}$ and $\phi_{z}$ for $J=1,\lambda=0.5,V=2,k=\frac{\pi}{2}$, (a) $\Delta=0$ with nodal points at quasienergy zero and (b) $\Delta=0.3$ where the nodal points evolve into nodal lines. Nodal lines with representative values of $\Delta$, (c) at quasienergy zero and (d) at quasienergy $\pi$.}
  \label{Fig3}
\end{figure}

The similarity and difference between the static and the kicked model at small and large values of parameters, respectively, can be understood physically as follows. By Fourier decomposing the kicked system in terms of its frequency components, it consists of the static version of the model, which is coupled with infinitely many frequency modes. If the system parameters are small, the transition frequency of the static version will be small (off-resonant) as compared with any of the frequency modes. As a result, the band structure should not change much, and the properties of the static version can be carried forward to the kicked version. As the system parameters increase, the transition frequency will in general increase. Consequently, at some values of $k$, $\phi_y$, and $\phi_z$, the transition frequency may become resonant with one of the frequency components of the driving (kicking) term. As a result, the two energy bands are dynamically connected, which leads to the different features of the kicked model, such as the emergence of additional band touching points at quasienergy $\pi$, which we showed earlier.

\subsection{$Z_{2}\times Z_{2}$ invariant}
In order to discuss the symmetry class of the time-evolution operator, we rewrite the Floquet operator with $T=1$ in Eq.~(11) in a symmetrized form~\cite{PhysRevB.88.155133},
\begin{eqnarray}
\label{eqn.19}
U(k,\phi_{y},\phi_{z})=e^{-i\frac{h_{1}}{2}}e^{-ih_{0}}e^{-i\frac{h_{1}}{2}}.
\end{eqnarray}
The Floquet operators in Eqs.~(11) and (19) have the same eigenvalues, with their eigenvectors related by a unitary transformation. The time-dependent Hamiltonian now has PHS, TRS, and CS~\cite{Roy2016Floquet},
\begin{eqnarray}
\label{eqn.20}
Ph(k,\phi_{y},\phi_{z},t)P^{-1}&=&-h(-k,-\phi_{y},-\phi_{z},t) \nonumber\\ Th(k,\phi_{y},\phi_{z},t)T^{-1}&=&h(-k,-\phi_{y},-\phi_{z},-t) \nonumber\\ C_{1}h(k,\phi_{y},\phi_{z},t)C^{-1}_{1}&=&-h(k,\phi_{y},\phi_{z},-t)
\end{eqnarray}
where $P$, $T$, and $C_1$ operators are the same as those in the static case. It follows that the time-evolution operator $U(k,\phi_{y},\phi_{z},t)$ for arbitrary $t\in[0,1]$ satisfies~\cite{Roy2016Floquet, 1367-2630-17-12-125014, PhysRevB.90.125143}
\begin{eqnarray}
\label{eqn.21}
PU(k,\phi_{y},\phi_{z},t)P^{-1}&=&U^{*}(-k,-\phi_{y},-\phi_{z},t) \nonumber\\ TU(k,\phi_{y},\phi_{z},t)T^{-1}&=&U^{*}(-k,-\phi_{y},-\phi_{z},1-t) \nonumber\\ C_{1}U(k,\phi_{y},\phi_{z},t)C^{-1}_{1}&=&U(k,\phi_{y},\phi_{z},1-t).
\end{eqnarray}
As a result, the time-evolution unitary operator falls into the time-dependant BDI class with no topological invariants in 3D and a $Z\times Z$ \footnote{The sign $\times$ here and in the $Z_{2}\times Z_{2}$ index represents a Cartesian product. The elements of $Z\times Z$ or $Z_{2}\times Z_{2}$ thus take the form of an ordered pair ($Z1$,$Z2$), where $Z1$ and $Z2$ are independent of each other.} index in 1D when $\phi_{y}$ and $\phi_{z}$ are treated as parameters~\cite{Roy2016Floquet}. In Floquet (topological) superconducting systems, two flavors of Floquet Majorana modes may emerge, i.e., one at quasienergy zero, the other at $\pi$, because both zero and $\pi$ are particle-hole symmetric due to the periodic nature of the quasienegy~\cite{PhysRevLett.106.220402, 1367-2630-17-12-125014}. For every 1D chain with $\phi_{y}$ and $\phi_{z}$ seen as parameters, the number of Floquet Majorana zero-mode $n_0$ and that of $\pi$-mode $n_{\pi}$ can constitute a $Z\times Z$ index, with their parities represented by two $Z_{2}$ indices, $Q_{0}=(-1)^{n_{0}}$, and $Q_{\pi}=(-1)^{n_{\pi}}$, respectively~\cite{PhysRevB.90.125143, 1367-2630-17-12-125014}. The two $Z_{2}$ indices, $Q_{0}$ and $Q_{\pi}$, constitute the $Z_{2}\times Z_{2}$ index introduced in Ref.~\cite{PhysRevLett.106.220402}, where $Q_{0}$ ($Q_{\pi}$) is defined by the parity of the total number of times that the eigenvalues of $U_{0}(\tau)$ and $U_{\pi/2}(\tau)$ cross $1$ ($-1$)~\cite{PhysRevLett.106.220402}. $Q_{0}$ and $Q_{\pi}$ are in general independent of each other and they together characterize the topological property of a Floquet system with PHS~\cite{PhysRevLett.106.220402, 1367-2630-17-12-125014}. For the purpose of this paper, only the $Z_{2}\times Z_{2}$ index is numerically calculated by the methods in Ref.~\cite{PhysRevLett.106.220402} (see Appendix C), which is enough to characterize the existence of Floquet Majorana flat bands, since at most one Majorana zero mode or $\pi$ mode is found for the parameters we choose in this paper. Consequently, a nontrivial $Q_{0}$ ($Q_{\pi}$) indicates the appearance of one Floquet Majorana zero ($\pi$) mode.

\begin{figure*}
  \centering
  \includegraphics[scale=0.85]{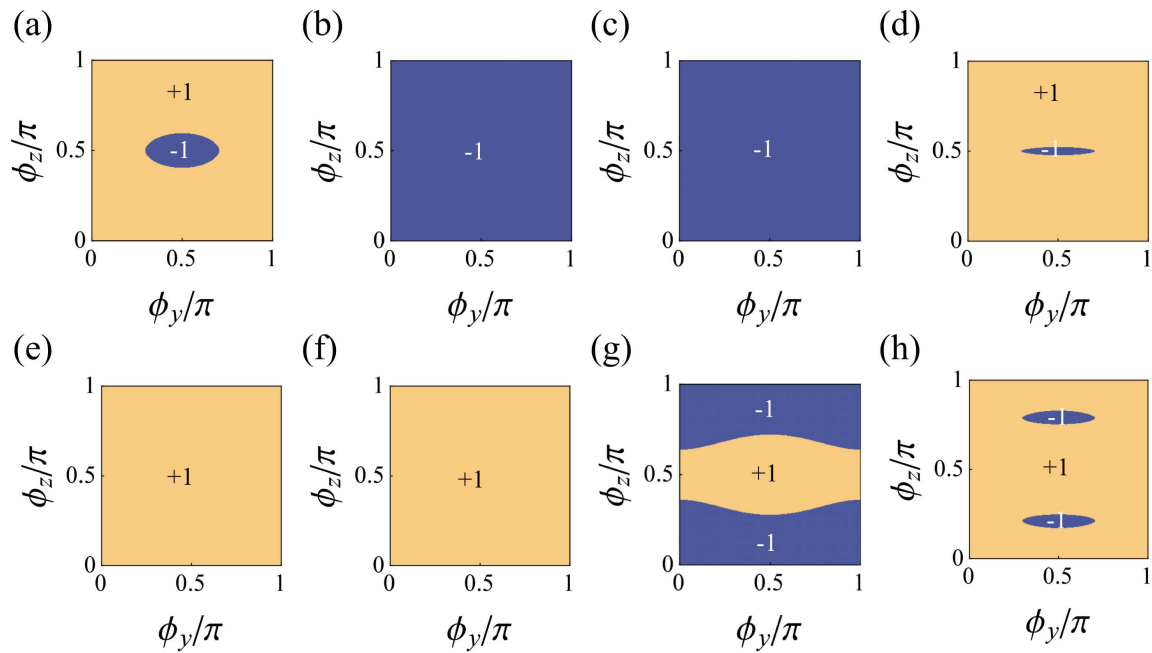}
  \caption{(Color online) The $Z_{2}\times Z_{2}$ index, $Q_{0}$ (upper row) and $Q_{\pi}$ (lower row) as a function of $\phi_{y}/\pi$ and $\phi_{z}/\pi$, for fixed $J=1,\lambda=0.5$, with $V=2,\Delta=0.3$ (first column), $V=2,\Delta=1.5$ (second column), $V=2,\Delta=2.5$ (third column), and $V=8,\Delta=0.3$ (last column). The $Q_{0,\pi}$ index inside (outside) the blue regions equals $-1(1)$, which indicates a nontrivial (trivial) phase.}
  \label{Fig4}
\end{figure*}

Results of the $Z_{2}\times Z_{2}$ index for four representative phases are presented in Fig.~4, with $Q_{0}$ ($Q_{\pi}$) values in the upper (lower) row, where $Q_{0}=-1$ ($Q_{\pi}=-1$) means nontrivial and is colored in blue. In the first column, the parameters are chosen as $J=1,\lambda=0.5,V=2,\Delta=0.3$, where the bulk Floquet spectrum exhibits nodal lines at quasienergy zero. It is shown that $Q_{0}=-1(1)$ inside (outside) the nodal lines, while $Q_{\pi}=1$ all over the region. In the last column with the parameters $J=1,\lambda=0.5,V=8,\Delta=0.3$, nodal loops appear at both quasienergy zero and $\pi$, and it is shown that $Q_{0}=-1(1)$ and $Q_{\pi}=-1(1)$ inside(outside) the nodal loops at zero and $\pi$, respectively. Of particular interest is the case where at least one of $Q_{0}$ and $Q_{\pi}$ is nontrivial for the whole $(\phi_{y},\phi_{z})$ region. As shown in the second column with $J=1,\lambda=0.5,V=2$ and $\Delta=1.5$, $Q_{0}$ always equals $-1$ in accordance with the absence of nodal lines at quasienergy zero. Similar to its corresponding static model as elucidated in Sec. II B, by bulk-edge correspondence, this implies the existence of zero mode throughout the whole parameter BZ. However, unlike the static model, by increasing $\Delta$ further, additional band touching points may appear at quasienergy $\pi$, leading to a region with nontrivial topology at quasienergy $\pi$, as illustrated in the third column of Fig.~4 with $J=1,\lambda=0.5,V=2$, and $\Delta=2.5$. This in turn leads to a region in which zero and $\pi$ modes emerge at the same $\phi_y$ and $\phi_z$, which will be verified directly in Fig.~5(e) in Sec. III C.

\subsection{ Floquet surface Majorana flat bands and Dirac arcs under OBCs}
In a 1D Floquet topological superconductor under OBCs, $Q_{0}=-1$ ($Q_{\pi}=-1$) indicates the emergence of Floquet Majorana modes at quasienergy zero ($\pi$)~\cite{PhysRevLett.106.220402}, in contrast to the static case where Majorana modes exist only at zero energy. Consequently, surface Majorana flat bands may emerge inside the projections of the nodal lines at both quasienergy zero and $\pi$. Figure 5 shows the quasienergy spectrum for fixed $J=1,\lambda=0.5$ and different parameters of $V$ and $\Delta$, where OBCs are taken along the lattice dimension with $N=60$ lattice sites. For $V=8$ and $\Delta=0.3$, the uppermost band near $\epsilon=\pi$ is shown in Fig.~5(a), where $\pi$-mode Floquet Majorana flat bands and arcs connecting them can be clearly seen. To verify their localized nature at the surface, we also choose one state, $(\phi_{y},\phi_{z})=(\frac{\pi}{2},\arccos\frac{\pi}{4})$, inside the $\pi$-mode flat band and plot the modular square of its particle component against the lattice sites $|u_{i}|^2$ in the inset of Fig.~5(a), which decays fast into the bulk. The quasienergy spectrum as a function of $\phi_{z}$ for fixed $\phi_{y}=\frac{\pi}{2}$ is shown in Fig.~5(b), where both zero-mode and $\pi$-mode Floquet Majorana flat bands can be observed.

At certain values of the system parameters, there exist arcs connecting different nodal loops at quasienergy zero or $\pi$, which can be seen as the remnants of the original Fermi arcs connecting different point nodes in Ref.~\cite{PhysRevE.93.022209} and also have a SSH-like origin. For example, in Fig.~5(a) with $l=1$ and $V=8$, the arcs are located around $\phi_{z}=\arccos\frac{\pi}{4}\approx0.212$ at quasienergy $\pi$. The emergence of these line modes can also be understood as follows. When $\Delta=0$, counterpropagating chiral edge states, whose origin has been elucidated in Sec. II C, exist at the same edge when $\phi_z$ is scanned from $-\pi$ to $\pi$ at a fixed $\phi_y$ between $-\frac{\pi}{2}$ and $\frac{\pi}{2}$, leading to a zero winding number inside each Floquet gap at quasienergy zero or $\pi$, and a zero Chern number for each Floquet band~\cite{PhysRevX.3.031005}, which is numerically shown in Fig.~5(c) with the parameters chosen as $J=1,\lambda=0.5,\Delta=0.3,\phi_{y}=0$, $V=8$, $\Delta=0$. After turning on a small $\Delta\neq0$, as long as the Floquet gap does not close and can be adiabatically connected to the $\Delta=0$ case, these chiral edge states will remain, as shown in Fig.~5(d) with $\Delta=0.3$. The crossings of these chiral edge states at quasienergy zero and $\pi$ form the line modes.

Moreover, as a feature of the kicked model, it is possible to generate Floquet Majorana modes at both quasienergy zero and $\pi$ simultaneously under the same $\phi_y$ and $\phi_z$. This can be achieved for example under $V=2, \Delta=2.5$ and $V=8,\Delta=2.2$, which are shown numerically in Figs.~5(e) and 5(f), respectively, where the quasienergy spectrum is plotted at a constant $\phi_y=\frac{\pi}{2}$ as a function of $\phi_z$. Figure 5(e) also numerically verifies the $Z_{2}\times Z_{2}$ phase diagrams in Figs.~4(c) and 4(g) in Sec. III B, where a large value of $\Delta$ gives rise to $\pi$-mode flat bands in addition to the zero-mode flat bands spanning the whole BZ. In this way, both of the flavors of Floquet Majorana modes as well as the region of the surface flat bands can be manipulated simultaneously by tuning the pairing order parameter.

\begin{figure*}
\centering
  \includegraphics[scale=1]{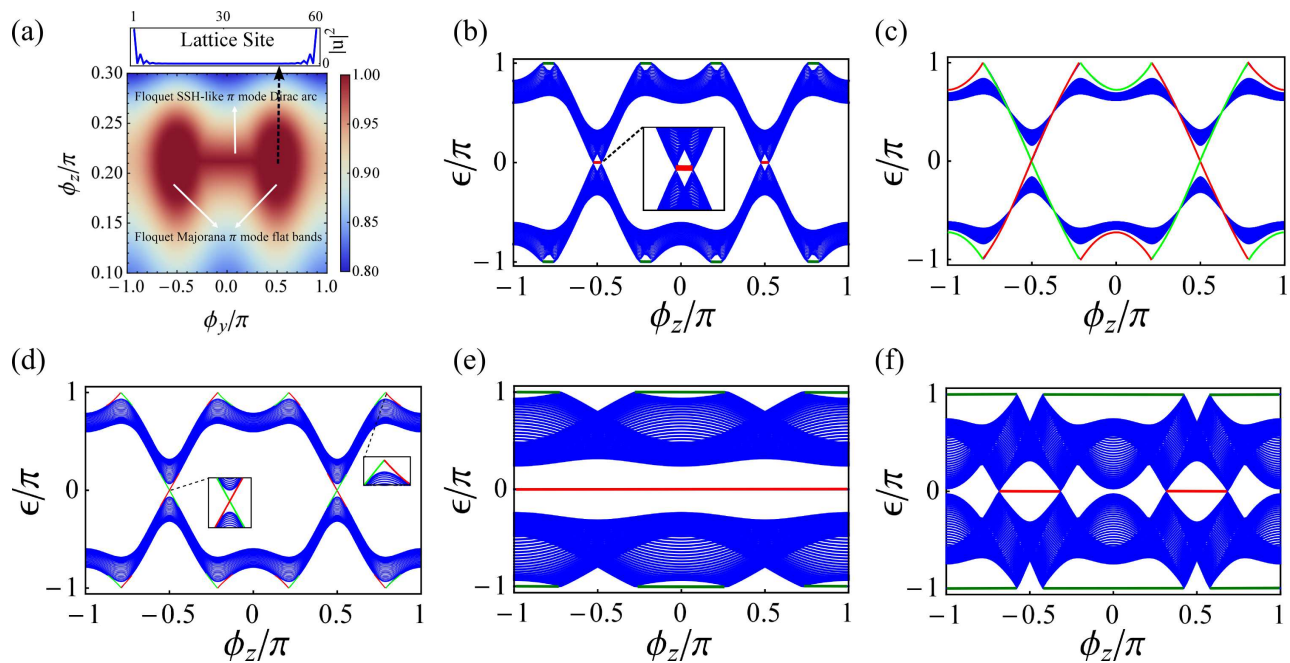}
\caption{(Color online) Quasienergy spectrum for fixed $J=1,\lambda=0.5$ and different $V,\Delta$ under OBCs in the lattice dimension with 60 lattice sites. For $V=8$ and $\Delta=0.3$, (a) the uppermost band around quasienergy $\pi$ as a function of $\phi_{y}$ and $\phi_{z}$, where Floquet Majorana $\pi$-mode flat bands as well as Floquet Dirac $\pi$-mode arcs connecting them can be seen. Inset: the modular square of the particle component of the wave function against lattice sites for $(\phi_{y},\phi_{z})=\big(\frac{\pi}{2},\arccos\frac{\pi}{4}\big)$ inside the $\pi$-mode flat bands. (b) The quasienergy spectrum as a function of $\phi_{z}$ for fixed $\phi_{y}=\frac{\pi}{2}$. (c) The quasienergy spectrum for $V=8$ and $\Delta=0$ as a function of $\phi_{z}$ with fixed $\phi_{y}=0$, where counterpropagating chiral edge states appear in both Floquet gaps near $\epsilon=0$ and $\epsilon=\pi$, which still exist in (d) with a small $\Delta=0.3$. (e) For $V=2,\Delta=2.5$ and (f) $V=8,\Delta=2.2$, the quasienergy spectrum as a function of $\phi_{z}$ for fixed $\phi_{y}=\frac{\pi}{2}$, where edge modes emerge at quasienergy zero and $\pi$ simultaneously under the same $\phi_y$ and $\phi_z$.}
\label{Fig5}
\end{figure*}
\section{Conclusion and Discussion}
We have studied both the static and periodically kicked off-diagonal Harper model with nearest-neighbor $p$-wave  superconducting pairing. It is found that, depending on the pairing order parameter, the particle bands and hole bands can touch at isolated points, along nodal lines, or even do not touch at all. In the static model, they touch at zero energy and the location of the nodal lines can be determined analytically. In the kicked model, they touch at quasienergy zero or $\pi$. By treating the two phase shifts as parameters, the topological property of each of the effective 1D chains along the lattice dimension is revealed by calculating corresponding topological invariants. The nodal lines correspond to boundaries between different topological phases since the change of topological invariant usually requires the closing-reopening process of the bulk (quasi)-energy gap. For nontrivial phases, when OBCs are taken along the lattice dimension, Majorana flat bands will appear in the 2D surface spanned by the parameters, which is bounded by the projections of nodal lines onto this surface. These flat bands are controllable and provide a large density of states for Floquet Majorana fermions and may help us study various properties of (Floquet) Majorana fermions. For certain parameter regions, there also exists (Floquet) SSH-like Dirac edge modes, which form arcs connecting different surface Majorana flat bands. This implies that both the (Floquet) SSH-like modes and (Floquet) Majorana modes can be realized in the same system. Despite the similarity between the static and kicked model, we have also emphasized a feature of the kicked model, i.e., the possibility to generate two controllable Majorana flat bands, one at quasienergy zero and the other at quasienergy $\pi$, at the same values of $\phi_y$ and $\phi_z$. This suggests that while the superconductivity term leads to novel physics in both the static and kicked model, the kicked model (or Floquet systems in general) might offer a more controllable platform for potential applications in the area of quantum computation and quantum control.

As a possible experimental realization of our model, we note that the $p$-wave pairing term can be introduced in the cold-atom setups through the $p$-wave Feshbach resonance~\cite{PhysRevLett.90.053201}, synthetic spin-orbit coupling~\cite{PhysRevLett.101.160401}, or by combining orbital degrees of freedom with strong $s$-wave interactions~\cite{B2014Majorana}. Moreover, the Harper model with artificial dimensions can be easily realized in either waveguide setups~\cite{PhysRevLett.110.076403} or cold-atom setups~\cite{PhysRevLett.112.043001}. By focusing on the cold atom realization of the Harper model in an optical lattice, the kicking term can be simulated by considering a two-step protocol as follows. For $nT<t\leq nT+\frac{T}{2}$, with $n$ and $T$ being an integer and the period of the protocol, respectively, the lattice depth can be tuned to be very small so that the hopping term between two lattice sites is dominant over the on-site potential term. For $nT+\frac{T}{2}<t\leq (n+1)T$, the lattice depth is then tuned to be very deep so that the hopping strength is minimized. This two-step protocol can be achieved by controlling the two counterpropagating lasers making up the optical lattice. Finally, by designing a mechanism that combines the realizations of $p$-wave pairing term and extended Harper model elucidated above, our model can be experimentally realized.

So far, we have only focused on the generalized Harper model with $\pi$ flux, i.e., $\alpha_1=\alpha_2=\frac{1}{2}$ in Eq.~(1), which possesses two bands (four bands in the presence of the $p$-wave pairing term). It is expected that considering the case with more than two bands might be more fruitful~\cite{0953-8984-29-3-035601}. Moreover, when next-nearest-neighbor coupling terms are introduced, the sublattice symmetry $C_{2}$ is broken since next-nearest neighbors belong to the same sublattice. Nevertheless, when only a real next-nearest-neighbor $p$-wave paring potential is considered, the SSH-like edge modes survive but they are no longer degenerate with a small energy splitting from each other. As for the chiral symmetry $C_{1}$, it may or may not remain, depending on whether TRS is broken by terms such as the phase difference between nearest-neighbor pairing potential and next-nearest-neighbor pairing potential. If this chiral symmetry is preserved, more than one Majorana zero (or $\pi$) mode may emerge on each 1D chain's end due to the longer-ranged couplings~\cite{PhysRevB.85.035110, PhysRevB.87.201109}, whose detailed analysis is beyond the scope of this paper. In addition, we have only considered the effect of the $p$-wave pairing term applied along one dimension, i.e., the physical dimension, in this work. In order to better simulate a physical 3D system, introducing the $p$-wave term along the other (artificial) dimensions is preferred. This can be accomplished by designing a scheme that simulates superconductivity along an artificial dimension, which remains an open question. Nevertheless, our results add knowledge on the effect of superconductivity on Weyl points and its implications on static and Floquet systems~\cite{PhysRevB.86.054504, PhysRevB.86.214514, PhysRevB.92.035153, PhysRevB.93.094517, PhysRevB.93.184511}. One possible future work along this direction might be to consider a more realistic Floquet system, such as the continuously driven Harper model ~\cite{Zhou2014} with off-diagonal modulation, which amounts to replacing the kicking term in Eq.~(9) with a harmonic driving. It is expected that this feature of the kicked model might also be found in such a model. Another topic that can be explored includes the relationship between the topological invariant in our model and transport properties, similar to the charge pumping~\cite{PhysRevB.27.6083, PhysRevLett.109.010601, PhysRevE.93.022209} or the fermion parity pumping~\cite{PhysRevLett.111.116402} over one adiabatic cycle. It is also interesting to investigate physical phenomena resulting from the coexistence of SSH-like Dirac modes and Kitaev-like Majorana modes. Finally, the emergence of controllable flat bands at quasienergy zero and $\pi$ might serve as a good starting point to study interaction effects in Floquet systems.

\textit{Note added.} Recently, we became aware of a paper~\cite{Zeng2016} which investigated the static $p$-wave superconducting pairing Harper model in both incommensurate and commensurate cases. Similar results for the commensurate cases of $\alpha_{1}=\alpha_{2}=\frac{1}{2}$ were found there. Both Kitaev-like Majorana zero modes and SSH-like Dirac zero modes are also shown in our paper.

H.Q.W., M.N.Chen., and R.W.B. contributed equally to this work.

\appendix
\section{Calculation of the $Z$ invariant}
We first calculate the chiral index $N_{1}$ corresponding to the chiral symmetry $C_{1}=\tau_{x}$ for general $\phi_{y}$ and $\phi_{z}$. By introducing a unitary transformation,
\begin{eqnarray}
\label{eqn.A1}
U_{1}=\frac{1}{\sqrt{2}}\left(
  \begin{array}{cccc}
    1 & 0 & 1 & 0 \\
    0 & 1 & 0 & 1 \\
    -1 & 0 & 1 & 0 \\
    0 & -1 & 0 & 1 \\
  \end{array}
\right)
\end{eqnarray}
which diagonalizes $C_{1}$ as $U_{1}C_{1}U_{1}^{\dagger}=\tau_{z}$, the Hamiltonian can be brought into an off-diagonal form,
\begin{eqnarray}
\label{eqn.A2}
U_{1}hU_{1}^{\dagger}=\left(
                        \begin{array}{cc}
                          0 & D_{1} \\
                           D_{1}^{\dagger}& 0 \\
                        \end{array}
                      \right),
\end{eqnarray}
with
\begin{eqnarray}
\label{eqn.A3}
D_{1}=\left(
    \begin{array}{cc}
      -c & -a+ib_{+} \\
      -a+ib_{-} & c \\
    \end{array}
  \right),
\end{eqnarray}
where $a=J\cos k$, $b_{\pm}=(\Delta\pm\lambda\cos\phi_{y})\sin k$, and $c=\frac{V\cos\phi_{z}}{2}$. The chiral index $N_{1}$ can then be calculated by~\cite{PhysRevB.90.014505,Zeng2016}
\begin{eqnarray}
\label{eqn.A4}
N_{1}=-\mathrm{Tr}\int_{0}^{\pi}\frac{dk}{2\pi i}D_{1}^{-1}\partial_{k}D_{1}
=-\int_{0}^{\pi}\frac{dk}{2\pi i}\partial_{k}\ln Z_{1},\nonumber\\
\end{eqnarray}
with
\begin{eqnarray}
\label{eqn.A5}
Z_{1}=\mathrm{Det} D_{1}=&&\Delta^{2}\sin^{2}k-J^2\cos^{2}k-\lambda^{2}\cos^{2}\phi_{y}\sin^{2}k\nonumber\\
&&-\frac{1}{4}V^{2}\cos^{2}\phi_{z}+2i \Delta J\sin k\cos k.
\end{eqnarray}
Here, $\Delta>0$, so we have
\begin{eqnarray}
\label{eqn.A6}
N_{1}=\left\{ \begin{array}{ll}
1& \textrm{for $\Delta>\sqrt{\lambda^{2}\cos^{2}\phi_{y}+\frac{V^{2}\cos^{2}\phi_{z}}{4}}$},\\
0 & \textrm{for $\Delta<\sqrt{\lambda^{2}\cos^{2}\phi_{y}+\frac{V^{2}\cos^{2}\phi_{z}}{4}}$}.
\end{array}\right.
\end{eqnarray}

Next we calculate the chiral index $N_{2}$ corresponding to the chiral (sublattice) symmetry $C_{2}=\sigma_{z}$ when $\phi_{z}=\pm\frac{\pi}{2}$. Similarly, we introduce a unitary transformation,
\begin{eqnarray}
\label{eqn.A7}
U_{2}=\left(
  \begin{array}{cccc}
    1 & 0 & 0 & 0 \\
    0 & 0 & 1 & 0 \\
    0 & 1 & 0 & 0 \\
    0 & 0 & 0 & 1 \\
  \end{array}
\right),
\end{eqnarray}
through which we get $U_{2}C_{2}U_{2}^{\dagger}=\tau_{z}$ and
\begin{eqnarray}
\label{eqn.A8}
U_{2}hU_{2}^{\dagger}=\left(
                        \begin{array}{cc}
                          0 & D_{2} \\
                           D_{2}^{\dagger}& 0 \\
                        \end{array}
                      \right),
\end{eqnarray}
with
\begin{eqnarray}
\label{eqn.A9}
D_{2}=\left(
    \begin{array}{cc}
     f & g \\
      -g & -f \\
    \end{array}
  \right),
\end{eqnarray}
where $f=J\cos k -i\lambda\cos\phi_{y}\sin k$ and $g=i \Delta\sin k$. To get the correct chiral index, a gauge transformation in the original Fourier transform $\hat{c}_{2,k}\rightarrow \hat{c}_{2,k}e^{-ik}$ needs to be performed to eliminate the phase difference between the two sublattices inside a unit cell. Under this transformation, $f\rightarrow fe^{ik}$ and $g\rightarrow ge^{ik}$. The chiral index $N_{2}$ can then be calculated as~\cite{PhysRevB.90.014505,Zeng2016}
\begin{eqnarray}
\label{eqn.A10}
N_{2}=-\mathrm{Tr}\int_{0}^{\pi}\frac{dk}{2\pi i}D_{2}^{-1}\partial_{k}D_{2}
=-\int_{0}^{\pi}\frac{dk}{2\pi i}\partial_{k}\ln Z_{2},\nonumber\\
\end{eqnarray}
with
\begin{eqnarray}
\label{eqn.A11}
Z_{2}=\mathrm{Det} D_{2}=&&\Big(\lambda^{2}\cos^{2}\phi_{y}\sin^{2}k-J^2\cos^{2}k-\Delta^{2}\sin^{2}k\nonumber\\
&&+2i \Delta J\sin k\cos k\Big)e^{2ik}.
\end{eqnarray}
Finally we get
\begin{eqnarray}
\label{eqn.A12}
N_{2}&=&\Theta(\lambda\cos\phi_{y}-\Delta)+\Theta(\lambda\cos\phi_{y}+\Delta)\nonumber\\
&=&\left\{ \begin{array}{ll}
2& \textrm{for $\lambda\cos\phi_{y}>\Delta$},\\
1& \textrm{for $-\Delta<\lambda\cos\phi_{y}<\Delta$},\\
0 & \textrm{for $\lambda\cos\phi_{y}<-\Delta$}.
\end{array}\right.
\end{eqnarray}
Here, $\Theta$ is the step function and $\Delta>0$.
\section{Calculation of the $Z_{2}$ invariant}
In order to calculate the $Z_{2}$ index, we need to rewrite the Hamiltonian in the Majorana representation through the transformation
\begin{eqnarray}
\label{eqn.B1}
\left(
  \begin{array}{c}
  \gamma_{1,k}\\
    \gamma_{2,k} \\
    \gamma_{3,k} \\
    \gamma_{4,k} \\
  \end{array}
\right)
=\left(
    \begin{array}{cccc}
      1 & 0 & 1 & 0 \\
      -i & 0 & i & 0 \\
      0 & 1 & 0 & 1 \\
      0 & -i & 0 & i \\
    \end{array}
  \right)
\left(
  \begin{array}{c}
  \hat{c}_{1,k}\\
    \hat{c}_{2,k} \\
    \hat{c}_{1,-k}^{\dagger} \\
    \hat{c}_{2,-k}^{\dagger} \\
  \end{array}
\right),
\end{eqnarray}
where the Majorana operators satisfy the following anti-commutation relation:
\begin{eqnarray}
\label{eqn.B2}
\{\gamma_{n,k},\gamma_{n',-k'}\}=2\delta_{nn'}\delta_{kk'}.\quad
(\gamma^{\dagger}_{n,k}=\gamma_{n,-k}).\nonumber\\
\end{eqnarray}
The Hamiltonian can be written in the form~\cite{Kitaev}
\begin{eqnarray}
\label{eqn.B3}
H=\frac{i}{4}\sum_{k}\sum_{nn'}M_{nn',k}\gamma_{n,-k}\gamma_{n',k},
\end{eqnarray}
with
\begin{eqnarray}
\label{eqn.B4}
M&=&2\left(
    \begin{array}{cccc}
       0&c  &0  &a+ib_{-}  \\
       -c& 0 &-a+ib_{+}   &0  \\
       0&a+ib_{+}  & 0 &-c  \\
       -a+ib_{-}&0  & c & 0 \\
    \end{array}
  \right),\nonumber\\
\end{eqnarray}
where $a,b_{\pm}$, and $c$ are the same as in Eq.~(A3). The $Z_{2}$ index is given by the sign of the products of the Pfaffian at two particle-hole symmetric momenta $k=0$ and $k=\frac{\pi}{2}$ \cite{Kitaev}
\begin{equation}
\label{eqn.B5}
  Q=\mathrm{sgn}\{\mathrm{Pf}[M_{0}]\mathrm{Pf}[M_{\pi/2}]\}
\end{equation}
where Pf means the Pfaffian of a matrix. However, in the above representation, $M_{\frac{\pi}{2}}$ is not skew symmetric, and its Pfaffian is not well defined. To make it skew symmetric, the same gauge transformation as in Appendix A is introduced to the Fourier transform $\hat{c}_{2,k}\rightarrow \hat{c}_{2,k}e^{-ik}$, leading to
\begin{eqnarray}
\label{eqn.B6}
M_{14}&\rightarrow&2(a+ib_{-})e^{-ik} \nonumber\\
M_{23}&\rightarrow&2(-a+ib_{+})e^{-ik}  \nonumber\\
M_{32}&\rightarrow&2(a+ib_{+})e^{ik} \nonumber\\
M_{41}&\rightarrow&2(-a+ib_{-})e^{ik}. \nonumber
\end{eqnarray}
The $Z_{2}$ index is then obtained as
\begin{equation}
\label{eqn.B7}
 Q=\mathrm{sgn}\bigg\{\bigg(\lambda^{2}\cos^{2}\phi_{y}+\frac{V^{2}\cos^{2}\phi_{z}}{4}\bigg)-\Delta^{2}\bigg\}.
\end{equation}
Here, $Q=-1$ stands for a nontrivial phase with odd number of Majorana zero modes, due to the relation $Q=(-1)^{N_{1}}$.
\section{Calculation of the $Z_{2}\times Z_{2}$ invariant}
In the Majorana representation as introduced in Appendix B, the time-independent term and the kicked term in the exponential of the Floquet operator can be expressed in the form~\cite{Kitaev}
\begin{eqnarray}
\label{eqn.C1}
H_{0}&=&\frac{i}{4}\sum_{k}\sum_{nn'}A_{nn',k}\gamma_{n,-k}\gamma_{n',k},\\
H_{\rm kicked}&=&\frac{i}{4}\sum_{k}\sum_{nn'}B_{nn',k}\gamma_{n,-k}\gamma_{n',k},
\end{eqnarray}
For the $A$ matrix, only the following four elements are nonzero:
\begin{eqnarray}
\label{eqn.C2}
A_{14}&=&(a+ib_{-})e^{-ik},\nonumber\\
A_{23}&=&(-a+ib_{+})e^{-ik},\nonumber\\
A_{32}&=&(a+ib_{+})e^{ik},\nonumber\\
A_{41}&=&(-a+ib_{-})e^{ik},\nonumber
\end{eqnarray}
where the same gauge transformation has been performed, while for the $B$ matrix,
\begin{eqnarray}
B&=&V\cos(\phi_{z})\delta(t-mT)\left(
    \begin{array}{cccc}
       0& 1 & 0 & 0 \\
       -1& 0 & 0 &0 \\
       0& 0 &0  &-1  \\
       0& 0 & 1 & 0 \\
    \end{array}
  \right),\nonumber
\end{eqnarray}
The $Z_{2}\times Z_{2}$ index can then be calculated by \cite{PhysRevLett.106.220402}
\begin{eqnarray}
  \label{eqn.C3}
  Q_{0}Q_{\pi}&=&\mathrm{sgn}\{\mathrm{Pf}[M_{0}]\mathrm{Pf}[M_{\pi/2}]\},\\ Q_{0}&=&\mathrm{sgn}\{\mathrm{Pf}[N_{0}]\mathrm{Pf}[N_{\pi/2}]\},
\end{eqnarray}
where $M_{k}=\ln[U_{k}]$ and $N_{k}=\ln[\sqrt{U_{k}}]$ are skew-symmetric matrices related to the evolution, and $\ln[\sqrt{U_{k}}]$ is derived from the analytic continuation from the history of $U_{k}(\tau)$.
\begin{acknowledgments}
We acknowledge Rui Wang for helpful discussions. This work was supported by the State Key Program for Basic Researches of China under Grant No. 2014CB921103. J.G. was funded by the Singapore Ministry of Education Academic Research Fund Tier 1 (WBS Grant No. R-144-000-353-112). H.Q.W. is supported by the program B for  Outstanding Ph.D. candidates of Nanjing University.
\end{acknowledgments}
\bibliography{Majorana}

\begin{thebibliography}{76}%
\makeatletter
\providecommand \@ifxundefined [1]{%
 \@ifx{#1\undefined}
}%
\providecommand \@ifnum [1]{%
 \ifnum #1\expandafter \@firstoftwo
 \else \expandafter \@secondoftwo
 \fi
}%
\providecommand \@ifx [1]{%
 \ifx #1\expandafter \@firstoftwo
 \else \expandafter \@secondoftwo
 \fi
}%
\providecommand \natexlab [1]{#1}%
\providecommand \enquote  [1]{``#1''}%
\providecommand \bibnamefont  [1]{#1}%
\providecommand \bibfnamefont [1]{#1}%
\providecommand \citenamefont [1]{#1}%
\providecommand \href@noop [0]{\@secondoftwo}%
\providecommand \href [0]{\begingroup \@sanitize@url \@href}%
\providecommand \@href[1]{\@@startlink{#1}\@@href}%
\providecommand \@@href[1]{\endgroup#1\@@endlink}%
\providecommand \@sanitize@url [0]{\catcode `\\12\catcode `\$12\catcode
  `\&12\catcode `\#12\catcode `\^12\catcode `\_12\catcode `\%12\relax}%
\providecommand \@@startlink[1]{}%
\providecommand \@@endlink[0]{}%
\providecommand \url  [0]{\begingroup\@sanitize@url \@url }%
\providecommand \@url [1]{\endgroup\@href {#1}{\urlprefix }}%
\providecommand \urlprefix  [0]{URL }%
\providecommand \Eprint [0]{\href }%
\providecommand \doibase [0]{http://dx.doi.org/}%
\providecommand \selectlanguage [0]{\@gobble}%
\providecommand \bibinfo  [0]{\@secondoftwo}%
\providecommand \bibfield  [0]{\@secondoftwo}%
\providecommand \translation [1]{[#1]}%
\providecommand \BibitemOpen [0]{}%
\providecommand \bibitemStop [0]{}%
\providecommand \bibitemNoStop [0]{.\EOS\space}%
\providecommand \EOS [0]{\spacefactor3000\relax}%
\providecommand \BibitemShut  [1]{\csname bibitem#1\endcsname}%
\let\auto@bib@innerbib\@empty
\bibitem [{\citenamefont {Hasan}\ and\ \citenamefont
  {Kane}(2010)}]{RevModPhys.82.3045}%
  \BibitemOpen
  \bibfield  {author} {\bibinfo {author} {\bibfnamefont {M.~Z.}\ \bibnamefont
  {Hasan}}\ and\ \bibinfo {author} {\bibfnamefont {C.~L.}\ \bibnamefont
  {Kane}},\ }\href {\doibase 10.1103/RevModPhys.82.3045} {\bibfield  {journal}
  {\bibinfo  {journal} {Rev. Mod. Phys.}\ }\textbf {\bibinfo {volume} {82}},\
  \bibinfo {pages} {3045} (\bibinfo {year} {2010})}\BibitemShut {NoStop}%
\bibitem [{\citenamefont {Qi}\ and\ \citenamefont
  {Zhang}(2011)}]{RevModPhys.83.1057}%
  \BibitemOpen
  \bibfield  {author} {\bibinfo {author} {\bibfnamefont {X.-L.}\ \bibnamefont
  {Qi}}\ and\ \bibinfo {author} {\bibfnamefont {S.-C.}\ \bibnamefont {Zhang}},\
  }\href {\doibase 10.1103/RevModPhys.83.1057} {\bibfield  {journal} {\bibinfo
  {journal} {Rev. Mod. Phys.}\ }\textbf {\bibinfo {volume} {83}},\ \bibinfo
  {pages} {1057} (\bibinfo {year} {2011})}\BibitemShut {NoStop}%
\bibitem [{\citenamefont {Wan}\ \emph {et~al.}(2011)\citenamefont {Wan},
  \citenamefont {Turner}, \citenamefont {Vishwanath},\ and\ \citenamefont
  {Savrasov}}]{PhysRevB.83.205101}%
  \BibitemOpen
  \bibfield  {author} {\bibinfo {author} {\bibfnamefont {X.}~\bibnamefont
  {Wan}}, \bibinfo {author} {\bibfnamefont {A.~M.}\ \bibnamefont {Turner}},
  \bibinfo {author} {\bibfnamefont {A.}~\bibnamefont {Vishwanath}}, \ and\
  \bibinfo {author} {\bibfnamefont {S.~Y.}\ \bibnamefont {Savrasov}},\ }\href
  {\doibase 10.1103/PhysRevB.83.205101} {\bibfield  {journal} {\bibinfo
  {journal} {Phys. Rev. B}\ }\textbf {\bibinfo {volume} {83}},\ \bibinfo
  {pages} {205101} (\bibinfo {year} {2011})}\BibitemShut {NoStop}%
\bibitem [{\citenamefont {Burkov}\ and\ \citenamefont
  {Balents}(2011)}]{PhysRevLett.107.127205}%
  \BibitemOpen
  \bibfield  {author} {\bibinfo {author} {\bibfnamefont {A.~A.}\ \bibnamefont
  {Burkov}}\ and\ \bibinfo {author} {\bibfnamefont {L.}~\bibnamefont
  {Balents}},\ }\href {\doibase 10.1103/PhysRevLett.107.127205} {\bibfield
  {journal} {\bibinfo  {journal} {Phys. Rev. Lett.}\ }\textbf {\bibinfo
  {volume} {107}},\ \bibinfo {pages} {127205} (\bibinfo {year}
  {2011})}\BibitemShut {NoStop}%
\bibitem [{\citenamefont {Xu}\ \emph {et~al.}(2015)\citenamefont {Xu},
  \citenamefont {Belopolski}, \citenamefont {Alidoust}, \citenamefont
  {Neupane}, \citenamefont {Bian}, \citenamefont {Zhang}, \citenamefont
  {Sankar}, \citenamefont {Chang}, \citenamefont {Yuan},\ and\ \citenamefont
  {Lee}}]{Xu2015TOPOLOGICAL}%
  \BibitemOpen
  \bibfield  {author} {\bibinfo {author} {\bibfnamefont {S.~Y.}\ \bibnamefont
  {Xu}}, \bibinfo {author} {\bibfnamefont {I.}~\bibnamefont {Belopolski}},
  \bibinfo {author} {\bibfnamefont {N.}~\bibnamefont {Alidoust}}, \bibinfo
  {author} {\bibfnamefont {M.}~\bibnamefont {Neupane}}, \bibinfo {author}
  {\bibfnamefont {G.}~\bibnamefont {Bian}}, \bibinfo {author} {\bibfnamefont
  {C.}~\bibnamefont {Zhang}}, \bibinfo {author} {\bibfnamefont
  {R.}~\bibnamefont {Sankar}}, \bibinfo {author} {\bibfnamefont
  {G.}~\bibnamefont {Chang}}, \bibinfo {author} {\bibfnamefont
  {Z.}~\bibnamefont {Yuan}}, \ and\ \bibinfo {author} {\bibfnamefont {C.~C.}\
  \bibnamefont {Lee}},\ }\href@noop {} {\bibfield  {journal} {\bibinfo
  {journal} {Science}\ }\textbf {\bibinfo {volume} {349}},\ \bibinfo {pages}
  {613} (\bibinfo {year} {2015})}\BibitemShut {NoStop}%
\bibitem [{\citenamefont {Weng}\ \emph {et~al.}(2015)\citenamefont {Weng},
  \citenamefont {Fang}, \citenamefont {Fang}, \citenamefont {Bernevig},\ and\
  \citenamefont {Dai}}]{PhysRevX.5.011029}%
  \BibitemOpen
  \bibfield  {author} {\bibinfo {author} {\bibfnamefont {H.}~\bibnamefont
  {Weng}}, \bibinfo {author} {\bibfnamefont {C.}~\bibnamefont {Fang}}, \bibinfo
  {author} {\bibfnamefont {Z.}~\bibnamefont {Fang}}, \bibinfo {author}
  {\bibfnamefont {B.~A.}\ \bibnamefont {Bernevig}}, \ and\ \bibinfo {author}
  {\bibfnamefont {X.}~\bibnamefont {Dai}},\ }\href {\doibase
  10.1103/PhysRevX.5.011029} {\bibfield  {journal} {\bibinfo  {journal} {Phys.
  Rev. X}\ }\textbf {\bibinfo {volume} {5}},\ \bibinfo {pages} {011029}
  (\bibinfo {year} {2015})}\BibitemShut {NoStop}%
\bibitem [{\citenamefont {Burkov}\ \emph {et~al.}(2011)\citenamefont {Burkov},
  \citenamefont {Hook},\ and\ \citenamefont {Balents}}]{PhysRevB.84.235126}%
  \BibitemOpen
  \bibfield  {author} {\bibinfo {author} {\bibfnamefont {A.~A.}\ \bibnamefont
  {Burkov}}, \bibinfo {author} {\bibfnamefont {M.~D.}\ \bibnamefont {Hook}}, \
  and\ \bibinfo {author} {\bibfnamefont {L.}~\bibnamefont {Balents}},\ }\href
  {\doibase 10.1103/PhysRevB.84.235126} {\bibfield  {journal} {\bibinfo
  {journal} {Phys. Rev. B}\ }\textbf {\bibinfo {volume} {84}},\ \bibinfo
  {pages} {235126} (\bibinfo {year} {2011})}\BibitemShut {NoStop}%
\bibitem [{\citenamefont {Phillips}\ and\ \citenamefont
  {Aji}(2014)}]{PhysRevB.90.115111}%
  \BibitemOpen
  \bibfield  {author} {\bibinfo {author} {\bibfnamefont {M.}~\bibnamefont
  {Phillips}}\ and\ \bibinfo {author} {\bibfnamefont {V.}~\bibnamefont {Aji}},\
  }\href {\doibase 10.1103/PhysRevB.90.115111} {\bibfield  {journal} {\bibinfo
  {journal} {Phys. Rev. B}\ }\textbf {\bibinfo {volume} {90}},\ \bibinfo
  {pages} {115111} (\bibinfo {year} {2014})}\BibitemShut {NoStop}%
\bibitem [{\citenamefont {Fang}\ \emph {et~al.}(2015)\citenamefont {Fang},
  \citenamefont {Chen}, \citenamefont {Kee},\ and\ \citenamefont
  {Fu}}]{PhysRevB.92.081201}%
  \BibitemOpen
  \bibfield  {author} {\bibinfo {author} {\bibfnamefont {C.}~\bibnamefont
  {Fang}}, \bibinfo {author} {\bibfnamefont {Y.}~\bibnamefont {Chen}}, \bibinfo
  {author} {\bibfnamefont {H.-Y.}\ \bibnamefont {Kee}}, \ and\ \bibinfo
  {author} {\bibfnamefont {L.}~\bibnamefont {Fu}},\ }\href {\doibase
  10.1103/PhysRevB.92.081201} {\bibfield  {journal} {\bibinfo  {journal} {Phys.
  Rev. B}\ }\textbf {\bibinfo {volume} {92}},\ \bibinfo {pages} {081201}
  (\bibinfo {year} {2015})}\BibitemShut {NoStop}%
\bibitem [{\citenamefont {Bian}\ \emph {et~al.}(2016)\citenamefont {Bian},
  \citenamefont {Chang}, \citenamefont {Zheng}, \citenamefont {Velury},
  \citenamefont {Xu}, \citenamefont {Neupert}, \citenamefont {Chiu},
  \citenamefont {Huang}, \citenamefont {Sanchez}, \citenamefont {Belopolski},
  \citenamefont {Alidoust}, \citenamefont {Chen}, \citenamefont {Chang},
  \citenamefont {Bansil}, \citenamefont {Jeng}, \citenamefont {Lin},\ and\
  \citenamefont {Hasan}}]{PhysRevB.93.121113}%
  \BibitemOpen
  \bibfield  {author} {\bibinfo {author} {\bibfnamefont {G.}~\bibnamefont
  {Bian}}, \bibinfo {author} {\bibfnamefont {T.-R.}\ \bibnamefont {Chang}},
  \bibinfo {author} {\bibfnamefont {H.}~\bibnamefont {Zheng}}, \bibinfo
  {author} {\bibfnamefont {S.}~\bibnamefont {Velury}}, \bibinfo {author}
  {\bibfnamefont {S.-Y.}\ \bibnamefont {Xu}}, \bibinfo {author} {\bibfnamefont
  {T.}~\bibnamefont {Neupert}}, \bibinfo {author} {\bibfnamefont {C.-K.}\
  \bibnamefont {Chiu}}, \bibinfo {author} {\bibfnamefont {S.-M.}\ \bibnamefont
  {Huang}}, \bibinfo {author} {\bibfnamefont {D.~S.}\ \bibnamefont {Sanchez}},
  \bibinfo {author} {\bibfnamefont {I.}~\bibnamefont {Belopolski}}, \bibinfo
  {author} {\bibfnamefont {N.}~\bibnamefont {Alidoust}}, \bibinfo {author}
  {\bibfnamefont {P.-J.}\ \bibnamefont {Chen}}, \bibinfo {author}
  {\bibfnamefont {G.}~\bibnamefont {Chang}}, \bibinfo {author} {\bibfnamefont
  {A.}~\bibnamefont {Bansil}}, \bibinfo {author} {\bibfnamefont {H.-T.}\
  \bibnamefont {Jeng}}, \bibinfo {author} {\bibfnamefont {H.}~\bibnamefont
  {Lin}}, \ and\ \bibinfo {author} {\bibfnamefont {M.~Z.}\ \bibnamefont
  {Hasan}},\ }\href {\doibase 10.1103/PhysRevB.93.121113} {\bibfield  {journal}
  {\bibinfo  {journal} {Phys. Rev. B}\ }\textbf {\bibinfo {volume} {93}},\
  \bibinfo {pages} {121113} (\bibinfo {year} {2016})}\BibitemShut {NoStop}%
\bibitem [{\citenamefont {Mullen}\ \emph {et~al.}(2015)\citenamefont {Mullen},
  \citenamefont {Uchoa},\ and\ \citenamefont
  {Glatzhofer}}]{PhysRevLett.115.026403}%
  \BibitemOpen
  \bibfield  {author} {\bibinfo {author} {\bibfnamefont {K.}~\bibnamefont
  {Mullen}}, \bibinfo {author} {\bibfnamefont {B.}~\bibnamefont {Uchoa}}, \
  and\ \bibinfo {author} {\bibfnamefont {D.~T.}\ \bibnamefont {Glatzhofer}},\
  }\href {\doibase 10.1103/PhysRevLett.115.026403} {\bibfield  {journal}
  {\bibinfo  {journal} {Phys. Rev. Lett.}\ }\textbf {\bibinfo {volume} {115}},\
  \bibinfo {pages} {026403} (\bibinfo {year} {2015})}\BibitemShut {NoStop}%
\bibitem [{\citenamefont {Kim}\ \emph {et~al.}(2015)\citenamefont {Kim},
  \citenamefont {Wieder}, \citenamefont {Kane},\ and\ \citenamefont
  {Rappe}}]{PhysRevLett.115.036806}%
  \BibitemOpen
  \bibfield  {author} {\bibinfo {author} {\bibfnamefont {Y.}~\bibnamefont
  {Kim}}, \bibinfo {author} {\bibfnamefont {B.~J.}\ \bibnamefont {Wieder}},
  \bibinfo {author} {\bibfnamefont {C.~L.}\ \bibnamefont {Kane}}, \ and\
  \bibinfo {author} {\bibfnamefont {A.~M.}\ \bibnamefont {Rappe}},\ }\href
  {\doibase 10.1103/PhysRevLett.115.036806} {\bibfield  {journal} {\bibinfo
  {journal} {Phys. Rev. Lett.}\ }\textbf {\bibinfo {volume} {115}},\ \bibinfo
  {pages} {036806} (\bibinfo {year} {2015})}\BibitemShut {NoStop}%
\bibitem [{\citenamefont {Yu}\ \emph {et~al.}(2015)\citenamefont {Yu},
  \citenamefont {Weng}, \citenamefont {Fang}, \citenamefont {Dai},\ and\
  \citenamefont {Hu}}]{PhysRevLett.115.036807}%
  \BibitemOpen
  \bibfield  {author} {\bibinfo {author} {\bibfnamefont {R.}~\bibnamefont
  {Yu}}, \bibinfo {author} {\bibfnamefont {H.}~\bibnamefont {Weng}}, \bibinfo
  {author} {\bibfnamefont {Z.}~\bibnamefont {Fang}}, \bibinfo {author}
  {\bibfnamefont {X.}~\bibnamefont {Dai}}, \ and\ \bibinfo {author}
  {\bibfnamefont {X.}~\bibnamefont {Hu}},\ }\href {\doibase
  10.1103/PhysRevLett.115.036807} {\bibfield  {journal} {\bibinfo  {journal}
  {Phys. Rev. Lett.}\ }\textbf {\bibinfo {volume} {115}},\ \bibinfo {pages}
  {036807} (\bibinfo {year} {2015})}\BibitemShut {NoStop}%
\bibitem [{\citenamefont {Schnyder}\ \emph {et~al.}(2008)\citenamefont
  {Schnyder}, \citenamefont {Ryu}, \citenamefont {Furusaki},\ and\
  \citenamefont {Ludwig}}]{PhysRevB.78.195125}%
  \BibitemOpen
  \bibfield  {author} {\bibinfo {author} {\bibfnamefont {A.~P.}\ \bibnamefont
  {Schnyder}}, \bibinfo {author} {\bibfnamefont {S.}~\bibnamefont {Ryu}},
  \bibinfo {author} {\bibfnamefont {A.}~\bibnamefont {Furusaki}}, \ and\
  \bibinfo {author} {\bibfnamefont {A.~W.~W.}\ \bibnamefont {Ludwig}},\ }\href
  {\doibase 10.1103/PhysRevB.78.195125} {\bibfield  {journal} {\bibinfo
  {journal} {Phys. Rev. B}\ }\textbf {\bibinfo {volume} {78}},\ \bibinfo
  {pages} {195125} (\bibinfo {year} {2008})}\BibitemShut {NoStop}%
\bibitem [{\citenamefont {Chiu}\ \emph {et~al.}(2016)\citenamefont {Chiu},
  \citenamefont {Teo}, \citenamefont {Schnyder},\ and\ \citenamefont
  {Ryu}}]{RevModPhys.88.035005}%
  \BibitemOpen
  \bibfield  {author} {\bibinfo {author} {\bibfnamefont {C.-K.}\ \bibnamefont
  {Chiu}}, \bibinfo {author} {\bibfnamefont {J.~C.~Y.}\ \bibnamefont {Teo}},
  \bibinfo {author} {\bibfnamefont {A.~P.}\ \bibnamefont {Schnyder}}, \ and\
  \bibinfo {author} {\bibfnamefont {S.}~\bibnamefont {Ryu}},\ }\href {\doibase
  10.1103/RevModPhys.88.035005} {\bibfield  {journal} {\bibinfo  {journal}
  {Rev. Mod. Phys.}\ }\textbf {\bibinfo {volume} {88}},\ \bibinfo {pages}
  {035005} (\bibinfo {year} {2016})}\BibitemShut {NoStop}%
\bibitem [{\citenamefont {Lindner}\ \emph {et~al.}(2010)\citenamefont
  {Lindner}, \citenamefont {Refael},\ and\ \citenamefont
  {Galitski}}]{Lindner2010Floquet}%
  \BibitemOpen
  \bibfield  {author} {\bibinfo {author} {\bibfnamefont {N.}~\bibnamefont
  {Lindner}}, \bibinfo {author} {\bibfnamefont {G.}~\bibnamefont {Refael}}, \
  and\ \bibinfo {author} {\bibfnamefont {V.}~\bibnamefont {Galitski}},\
  }\href@noop {} {\bibfield  {journal} {\bibinfo  {journal} {Nature Physics}\
  }\textbf {\bibinfo {volume} {7}},\ \bibinfo {pages} {490} (\bibinfo {year}
  {2010})}\BibitemShut {NoStop}%
\bibitem [{\citenamefont {Oka}\ and\ \citenamefont
  {Aoki}(2009)}]{PhysRevB.79.081406}%
  \BibitemOpen
  \bibfield  {author} {\bibinfo {author} {\bibfnamefont {T.}~\bibnamefont
  {Oka}}\ and\ \bibinfo {author} {\bibfnamefont {H.}~\bibnamefont {Aoki}},\
  }\href {\doibase 10.1103/PhysRevB.79.081406} {\bibfield  {journal} {\bibinfo
  {journal} {Phys. Rev. B}\ }\textbf {\bibinfo {volume} {79}},\ \bibinfo
  {pages} {081406} (\bibinfo {year} {2009})}\BibitemShut {NoStop}%
\bibitem [{\citenamefont {Kitagawa}\ \emph {et~al.}(2010)\citenamefont
  {Kitagawa}, \citenamefont {Berg}, \citenamefont {Rudner},\ and\ \citenamefont
  {Demler}}]{PhysRevB.82.235114}%
  \BibitemOpen
  \bibfield  {author} {\bibinfo {author} {\bibfnamefont {T.}~\bibnamefont
  {Kitagawa}}, \bibinfo {author} {\bibfnamefont {E.}~\bibnamefont {Berg}},
  \bibinfo {author} {\bibfnamefont {M.}~\bibnamefont {Rudner}}, \ and\ \bibinfo
  {author} {\bibfnamefont {E.}~\bibnamefont {Demler}},\ }\href {\doibase
  10.1103/PhysRevB.82.235114} {\bibfield  {journal} {\bibinfo  {journal} {Phys.
  Rev. B}\ }\textbf {\bibinfo {volume} {82}},\ \bibinfo {pages} {235114}
  (\bibinfo {year} {2010})}\BibitemShut {NoStop}%
\bibitem [{\citenamefont {Dahlhaus}\ \emph {et~al.}(2011)\citenamefont
  {Dahlhaus}, \citenamefont {Edge}, \citenamefont {Tworzyd\l{}o},\ and\
  \citenamefont {Beenakker}}]{PhysRevB.84.115133}%
  \BibitemOpen
  \bibfield  {author} {\bibinfo {author} {\bibfnamefont {J.~P.}\ \bibnamefont
  {Dahlhaus}}, \bibinfo {author} {\bibfnamefont {J.~M.}\ \bibnamefont {Edge}},
  \bibinfo {author} {\bibfnamefont {J.}~\bibnamefont {Tworzyd\l{}o}}, \ and\
  \bibinfo {author} {\bibfnamefont {C.~W.~J.}\ \bibnamefont {Beenakker}},\
  }\href {\doibase 10.1103/PhysRevB.84.115133} {\bibfield  {journal} {\bibinfo
  {journal} {Phys. Rev. B}\ }\textbf {\bibinfo {volume} {84}},\ \bibinfo
  {pages} {115133} (\bibinfo {year} {2011})}\BibitemShut {NoStop}%
\bibitem [{\citenamefont {Tong}\ \emph {et~al.}(2013)\citenamefont {Tong},
  \citenamefont {An}, \citenamefont {Gong}, \citenamefont {Luo},\ and\
  \citenamefont {Oh}}]{PhysRevB.87.201109}%
  \BibitemOpen
  \bibfield  {author} {\bibinfo {author} {\bibfnamefont {Q.-J.}\ \bibnamefont
  {Tong}}, \bibinfo {author} {\bibfnamefont {J.-H.}\ \bibnamefont {An}},
  \bibinfo {author} {\bibfnamefont {J.}~\bibnamefont {Gong}}, \bibinfo {author}
  {\bibfnamefont {H.-G.}\ \bibnamefont {Luo}}, \ and\ \bibinfo {author}
  {\bibfnamefont {C.~H.}\ \bibnamefont {Oh}},\ }\href {\doibase
  10.1103/PhysRevB.87.201109} {\bibfield  {journal} {\bibinfo  {journal} {Phys.
  Rev. B}\ }\textbf {\bibinfo {volume} {87}},\ \bibinfo {pages} {201109}
  (\bibinfo {year} {2013})}\BibitemShut {NoStop}%
\bibitem [{\citenamefont {Lindner}\ \emph {et~al.}(2013)\citenamefont
  {Lindner}, \citenamefont {Bergman}, \citenamefont {Refael},\ and\
  \citenamefont {Galitski}}]{PhysRevB.87.235131}%
  \BibitemOpen
  \bibfield  {author} {\bibinfo {author} {\bibfnamefont {N.~H.}\ \bibnamefont
  {Lindner}}, \bibinfo {author} {\bibfnamefont {D.~L.}\ \bibnamefont
  {Bergman}}, \bibinfo {author} {\bibfnamefont {G.}~\bibnamefont {Refael}}, \
  and\ \bibinfo {author} {\bibfnamefont {V.}~\bibnamefont {Galitski}},\ }\href
  {\doibase 10.1103/PhysRevB.87.235131} {\bibfield  {journal} {\bibinfo
  {journal} {Phys. Rev. B}\ }\textbf {\bibinfo {volume} {87}},\ \bibinfo
  {pages} {235131} (\bibinfo {year} {2013})}\BibitemShut {NoStop}%
\bibitem [{\citenamefont {Thakurathi}\ \emph {et~al.}(2013)\citenamefont
  {Thakurathi}, \citenamefont {Patel}, \citenamefont {Sen},\ and\ \citenamefont
  {Dutta}}]{PhysRevB.88.155133}%
  \BibitemOpen
  \bibfield  {author} {\bibinfo {author} {\bibfnamefont {M.}~\bibnamefont
  {Thakurathi}}, \bibinfo {author} {\bibfnamefont {A.~A.}\ \bibnamefont
  {Patel}}, \bibinfo {author} {\bibfnamefont {D.}~\bibnamefont {Sen}}, \ and\
  \bibinfo {author} {\bibfnamefont {A.}~\bibnamefont {Dutta}},\ }\href
  {\doibase 10.1103/PhysRevB.88.155133} {\bibfield  {journal} {\bibinfo
  {journal} {Phys. Rev. B}\ }\textbf {\bibinfo {volume} {88}},\ \bibinfo
  {pages} {155133} (\bibinfo {year} {2013})}\BibitemShut {NoStop}%
\bibitem [{\citenamefont {Wang}\ \emph {et~al.}(2015)\citenamefont {Wang},
  \citenamefont {Zhou},\ and\ \citenamefont {Gong}}]{PhysRevB.91.085420}%
  \BibitemOpen
  \bibfield  {author} {\bibinfo {author} {\bibfnamefont {H.}~\bibnamefont
  {Wang}}, \bibinfo {author} {\bibfnamefont {L.}~\bibnamefont {Zhou}}, \ and\
  \bibinfo {author} {\bibfnamefont {J.}~\bibnamefont {Gong}},\ }\href {\doibase
  10.1103/PhysRevB.91.085420} {\bibfield  {journal} {\bibinfo  {journal} {Phys.
  Rev. B}\ }\textbf {\bibinfo {volume} {91}},\ \bibinfo {pages} {085420}
  (\bibinfo {year} {2015})}\BibitemShut {NoStop}%
\bibitem [{\citenamefont {Ho}\ and\ \citenamefont
  {Gong}(2014)}]{PhysRevB.90.195419}%
  \BibitemOpen
  \bibfield  {author} {\bibinfo {author} {\bibfnamefont {D.~Y.~H.}\
  \bibnamefont {Ho}}\ and\ \bibinfo {author} {\bibfnamefont {J.}~\bibnamefont
  {Gong}},\ }\href {\doibase 10.1103/PhysRevB.90.195419} {\bibfield  {journal}
  {\bibinfo  {journal} {Phys. Rev. B}\ }\textbf {\bibinfo {volume} {90}},\
  \bibinfo {pages} {195419} (\bibinfo {year} {2014})}\BibitemShut {NoStop}%
\bibitem [{\citenamefont {Bomantara}\ \emph {et~al.}(2016)\citenamefont
  {Bomantara}, \citenamefont {Raghava}, \citenamefont {Zhou},\ and\
  \citenamefont {Gong}}]{PhysRevE.93.022209}%
  \BibitemOpen
  \bibfield  {author} {\bibinfo {author} {\bibfnamefont {R.~W.}\ \bibnamefont
  {Bomantara}}, \bibinfo {author} {\bibfnamefont {G.~N.}\ \bibnamefont
  {Raghava}}, \bibinfo {author} {\bibfnamefont {L.}~\bibnamefont {Zhou}}, \
  and\ \bibinfo {author} {\bibfnamefont {J.}~\bibnamefont {Gong}},\ }\href
  {\doibase 10.1103/PhysRevE.93.022209} {\bibfield  {journal} {\bibinfo
  {journal} {Phys. Rev. E}\ }\textbf {\bibinfo {volume} {93}},\ \bibinfo
  {pages} {022209} (\bibinfo {year} {2016})}\BibitemShut {NoStop}%
\bibitem [{\citenamefont {Inoue}\ and\ \citenamefont
  {Tanaka}(2010)}]{PhysRevLett.105.017401}%
  \BibitemOpen
  \bibfield  {author} {\bibinfo {author} {\bibfnamefont {J.-i.}\ \bibnamefont
  {Inoue}}\ and\ \bibinfo {author} {\bibfnamefont {A.}~\bibnamefont {Tanaka}},\
  }\href {\doibase 10.1103/PhysRevLett.105.017401} {\bibfield  {journal}
  {\bibinfo  {journal} {Phys. Rev. Lett.}\ }\textbf {\bibinfo {volume} {105}},\
  \bibinfo {pages} {017401} (\bibinfo {year} {2010})}\BibitemShut {NoStop}%
\bibitem [{\citenamefont {Jiang}\ \emph {et~al.}(2011)\citenamefont {Jiang},
  \citenamefont {Kitagawa}, \citenamefont {Alicea}, \citenamefont {Akhmerov},
  \citenamefont {Pekker}, \citenamefont {Refael}, \citenamefont {Cirac},
  \citenamefont {Demler}, \citenamefont {Lukin},\ and\ \citenamefont
  {Zoller}}]{PhysRevLett.106.220402}%
  \BibitemOpen
  \bibfield  {author} {\bibinfo {author} {\bibfnamefont {L.}~\bibnamefont
  {Jiang}}, \bibinfo {author} {\bibfnamefont {T.}~\bibnamefont {Kitagawa}},
  \bibinfo {author} {\bibfnamefont {J.}~\bibnamefont {Alicea}}, \bibinfo
  {author} {\bibfnamefont {A.~R.}\ \bibnamefont {Akhmerov}}, \bibinfo {author}
  {\bibfnamefont {D.}~\bibnamefont {Pekker}}, \bibinfo {author} {\bibfnamefont
  {G.}~\bibnamefont {Refael}}, \bibinfo {author} {\bibfnamefont {J.~I.}\
  \bibnamefont {Cirac}}, \bibinfo {author} {\bibfnamefont {E.}~\bibnamefont
  {Demler}}, \bibinfo {author} {\bibfnamefont {M.~D.}\ \bibnamefont {Lukin}}, \
  and\ \bibinfo {author} {\bibfnamefont {P.}~\bibnamefont {Zoller}},\ }\href
  {\doibase 10.1103/PhysRevLett.106.220402} {\bibfield  {journal} {\bibinfo
  {journal} {Phys. Rev. Lett.}\ }\textbf {\bibinfo {volume} {106}},\ \bibinfo
  {pages} {220402} (\bibinfo {year} {2011})}\BibitemShut {NoStop}%
\bibitem [{\citenamefont {Gu}\ \emph {et~al.}(2011)\citenamefont {Gu},
  \citenamefont {Fertig}, \citenamefont {Arovas},\ and\ \citenamefont
  {Auerbach}}]{PhysRevLett.107.216601}%
  \BibitemOpen
  \bibfield  {author} {\bibinfo {author} {\bibfnamefont {Z.}~\bibnamefont
  {Gu}}, \bibinfo {author} {\bibfnamefont {H.~A.}\ \bibnamefont {Fertig}},
  \bibinfo {author} {\bibfnamefont {D.~P.}\ \bibnamefont {Arovas}}, \ and\
  \bibinfo {author} {\bibfnamefont {A.}~\bibnamefont {Auerbach}},\ }\href
  {\doibase 10.1103/PhysRevLett.107.216601} {\bibfield  {journal} {\bibinfo
  {journal} {Phys. Rev. Lett.}\ }\textbf {\bibinfo {volume} {107}},\ \bibinfo
  {pages} {216601} (\bibinfo {year} {2011})}\BibitemShut {NoStop}%
\bibitem [{\citenamefont {D\'ora}\ \emph {et~al.}(2012)\citenamefont {D\'ora},
  \citenamefont {Cayssol}, \citenamefont {Simon},\ and\ \citenamefont
  {Moessner}}]{PhysRevLett.108.056602}%
  \BibitemOpen
  \bibfield  {author} {\bibinfo {author} {\bibfnamefont {B.}~\bibnamefont
  {D\'ora}}, \bibinfo {author} {\bibfnamefont {J.}~\bibnamefont {Cayssol}},
  \bibinfo {author} {\bibfnamefont {F.}~\bibnamefont {Simon}}, \ and\ \bibinfo
  {author} {\bibfnamefont {R.}~\bibnamefont {Moessner}},\ }\href {\doibase
  10.1103/PhysRevLett.108.056602} {\bibfield  {journal} {\bibinfo  {journal}
  {Phys. Rev. Lett.}\ }\textbf {\bibinfo {volume} {108}},\ \bibinfo {pages}
  {056602} (\bibinfo {year} {2012})}\BibitemShut {NoStop}%
\bibitem [{\citenamefont {Ezawa}(2013)}]{PhysRevLett.110.026603}%
  \BibitemOpen
  \bibfield  {author} {\bibinfo {author} {\bibfnamefont {M.}~\bibnamefont
  {Ezawa}},\ }\href {\doibase 10.1103/PhysRevLett.110.026603} {\bibfield
  {journal} {\bibinfo  {journal} {Phys. Rev. Lett.}\ }\textbf {\bibinfo
  {volume} {110}},\ \bibinfo {pages} {026603} (\bibinfo {year}
  {2013})}\BibitemShut {NoStop}%
\bibitem [{\citenamefont {G\'omez-Le\'on}\ and\ \citenamefont
  {Platero}(2013)}]{PhysRevLett.110.200403}%
  \BibitemOpen
  \bibfield  {author} {\bibinfo {author} {\bibfnamefont {A.}~\bibnamefont
  {G\'omez-Le\'on}}\ and\ \bibinfo {author} {\bibfnamefont {G.}~\bibnamefont
  {Platero}},\ }\href {\doibase 10.1103/PhysRevLett.110.200403} {\bibfield
  {journal} {\bibinfo  {journal} {Phys. Rev. Lett.}\ }\textbf {\bibinfo
  {volume} {110}},\ \bibinfo {pages} {200403} (\bibinfo {year}
  {2013})}\BibitemShut {NoStop}%
\bibitem [{\citenamefont {Liu}\ \emph {et~al.}(2013)\citenamefont {Liu},
  \citenamefont {Levchenko},\ and\ \citenamefont
  {Baranger}}]{PhysRevLett.111.047002}%
  \BibitemOpen
  \bibfield  {author} {\bibinfo {author} {\bibfnamefont {D.~E.}\ \bibnamefont
  {Liu}}, \bibinfo {author} {\bibfnamefont {A.}~\bibnamefont {Levchenko}}, \
  and\ \bibinfo {author} {\bibfnamefont {H.~U.}\ \bibnamefont {Baranger}},\
  }\href {\doibase 10.1103/PhysRevLett.111.047002} {\bibfield  {journal}
  {\bibinfo  {journal} {Phys. Rev. Lett.}\ }\textbf {\bibinfo {volume} {111}},\
  \bibinfo {pages} {047002} (\bibinfo {year} {2013})}\BibitemShut {NoStop}%
\bibitem [{\citenamefont {Lababidi}\ \emph {et~al.}(2014)\citenamefont
  {Lababidi}, \citenamefont {Satija},\ and\ \citenamefont
  {Zhao}}]{PhysRevLett.112.026805}%
  \BibitemOpen
  \bibfield  {author} {\bibinfo {author} {\bibfnamefont {M.}~\bibnamefont
  {Lababidi}}, \bibinfo {author} {\bibfnamefont {I.~I.}\ \bibnamefont
  {Satija}}, \ and\ \bibinfo {author} {\bibfnamefont {E.}~\bibnamefont
  {Zhao}},\ }\href {\doibase 10.1103/PhysRevLett.112.026805} {\bibfield
  {journal} {\bibinfo  {journal} {Phys. Rev. Lett.}\ }\textbf {\bibinfo
  {volume} {112}},\ \bibinfo {pages} {026805} (\bibinfo {year}
  {2014})}\BibitemShut {NoStop}%
\bibitem [{\citenamefont {Rudner}\ \emph {et~al.}(2013)\citenamefont {Rudner},
  \citenamefont {Lindner}, \citenamefont {Berg},\ and\ \citenamefont
  {Levin}}]{PhysRevX.3.031005}%
  \BibitemOpen
  \bibfield  {author} {\bibinfo {author} {\bibfnamefont {M.~S.}\ \bibnamefont
  {Rudner}}, \bibinfo {author} {\bibfnamefont {N.~H.}\ \bibnamefont {Lindner}},
  \bibinfo {author} {\bibfnamefont {E.}~\bibnamefont {Berg}}, \ and\ \bibinfo
  {author} {\bibfnamefont {M.}~\bibnamefont {Levin}},\ }\href {\doibase
  10.1103/PhysRevX.3.031005} {\bibfield  {journal} {\bibinfo  {journal} {Phys.
  Rev. X}\ }\textbf {\bibinfo {volume} {3}},\ \bibinfo {pages} {031005}
  (\bibinfo {year} {2013})}\BibitemShut {NoStop}%
\bibitem [{\citenamefont {Wang}\ \emph
  {et~al.}(2014{\natexlab{a}})\citenamefont {Wang}, \citenamefont {Wang},
  \citenamefont {Shen}, \citenamefont {Sheng},\ and\ \citenamefont
  {Xing}}]{0295-5075-105-1-17004}%
  \BibitemOpen
  \bibfield  {author} {\bibinfo {author} {\bibfnamefont {R.}~\bibnamefont
  {Wang}}, \bibinfo {author} {\bibfnamefont {B.}~\bibnamefont {Wang}}, \bibinfo
  {author} {\bibfnamefont {R.}~\bibnamefont {Shen}}, \bibinfo {author}
  {\bibfnamefont {L.}~\bibnamefont {Sheng}}, \ and\ \bibinfo {author}
  {\bibfnamefont {D.~Y.}\ \bibnamefont {Xing}},\ }\href
  {http://stacks.iop.org/0295-5075/105/i=1/a=17004} {\bibfield  {journal}
  {\bibinfo  {journal} {EPL (Europhysics Letters)}\ }\textbf {\bibinfo {volume}
  {105}},\ \bibinfo {pages} {17004} (\bibinfo {year}
  {2014}{\natexlab{a}})}\BibitemShut {NoStop}%
\bibitem [{\citenamefont {Klinovaja}\ \emph {et~al.}(2016)\citenamefont
  {Klinovaja}, \citenamefont {Stano},\ and\ \citenamefont
  {Loss}}]{PhysRevLett.116.176401}%
  \BibitemOpen
  \bibfield  {author} {\bibinfo {author} {\bibfnamefont {J.}~\bibnamefont
  {Klinovaja}}, \bibinfo {author} {\bibfnamefont {P.}~\bibnamefont {Stano}}, \
  and\ \bibinfo {author} {\bibfnamefont {D.}~\bibnamefont {Loss}},\ }\href
  {\doibase 10.1103/PhysRevLett.116.176401} {\bibfield  {journal} {\bibinfo
  {journal} {Phys. Rev. Lett.}\ }\textbf {\bibinfo {volume} {116}},\ \bibinfo
  {pages} {176401} (\bibinfo {year} {2016})}\BibitemShut {NoStop}%
\bibitem [{\citenamefont {Thakurathi}\ \emph {et~al.}(2016)\citenamefont
  {Thakurathi}, \citenamefont {Loss},\ and\ \citenamefont
  {Klinovaja}}]{Manisha2016Floquet}%
  \BibitemOpen
  \bibfield  {author} {\bibinfo {author} {\bibfnamefont {M.}~\bibnamefont
  {Thakurathi}}, \bibinfo {author} {\bibfnamefont {D.}~\bibnamefont {Loss}}, \
  and\ \bibinfo {author} {\bibfnamefont {J.}~\bibnamefont {Klinovaja}},\
  }\href@noop {} {\bibfield  {journal} {\bibinfo  {journal} {arXiv:1608.08143}\
  } (\bibinfo {year} {2016})}\BibitemShut {NoStop}%
\bibitem [{\citenamefont {Fulga}\ and\ \citenamefont
  {Maksymenko}(2016)}]{PhysRevB.93.075405}%
  \BibitemOpen
  \bibfield  {author} {\bibinfo {author} {\bibfnamefont {I.~C.}\ \bibnamefont
  {Fulga}}\ and\ \bibinfo {author} {\bibfnamefont {M.}~\bibnamefont
  {Maksymenko}},\ }\href {\doibase 10.1103/PhysRevB.93.075405} {\bibfield
  {journal} {\bibinfo  {journal} {Phys. Rev. B}\ }\textbf {\bibinfo {volume}
  {93}},\ \bibinfo {pages} {075405} (\bibinfo {year} {2016})}\BibitemShut
  {NoStop}%
\bibitem [{\citenamefont {Chen}\ \emph {et~al.}(2016)\citenamefont {Chen},
  \citenamefont {Su}, \citenamefont {Deng}, \citenamefont {Ruan}, \citenamefont
  {Luo}, \citenamefont {Shao}, \citenamefont {Sheng},\ and\ \citenamefont
  {Xing}}]{PhysRevB.94.205429}%
  \BibitemOpen
  \bibfield  {author} {\bibinfo {author} {\bibfnamefont {M.~N.}\ \bibnamefont
  {Chen}}, \bibinfo {author} {\bibfnamefont {W.}~\bibnamefont {Su}}, \bibinfo
  {author} {\bibfnamefont {M.~X.}\ \bibnamefont {Deng}}, \bibinfo {author}
  {\bibfnamefont {J.}~\bibnamefont {Ruan}}, \bibinfo {author} {\bibfnamefont
  {W.}~\bibnamefont {Luo}}, \bibinfo {author} {\bibfnamefont {D.~X.}\
  \bibnamefont {Shao}}, \bibinfo {author} {\bibfnamefont {L.}~\bibnamefont
  {Sheng}}, \ and\ \bibinfo {author} {\bibfnamefont {D.~Y.}\ \bibnamefont
  {Xing}},\ }\href {\doibase 10.1103/PhysRevB.94.205429} {\bibfield  {journal}
  {\bibinfo  {journal} {Phys. Rev. B}\ }\textbf {\bibinfo {volume} {94}},\
  \bibinfo {pages} {205429} (\bibinfo {year} {2016})}\BibitemShut {NoStop}%
\bibitem [{\citenamefont {Fruchart}(2016)}]{PhysRevB.93.115429}%
  \BibitemOpen
  \bibfield  {author} {\bibinfo {author} {\bibfnamefont {M.}~\bibnamefont
  {Fruchart}},\ }\href {\doibase 10.1103/PhysRevB.93.115429} {\bibfield
  {journal} {\bibinfo  {journal} {Phys. Rev. B}\ }\textbf {\bibinfo {volume}
  {93}},\ \bibinfo {pages} {115429} (\bibinfo {year} {2016})}\BibitemShut
  {NoStop}%
\bibitem [{\citenamefont {Nathan}\ and\ \citenamefont
  {Rudner}(2015)}]{1367-2630-17-12-125014}%
  \BibitemOpen
  \bibfield  {author} {\bibinfo {author} {\bibfnamefont {F.}~\bibnamefont
  {Nathan}}\ and\ \bibinfo {author} {\bibfnamefont {M.~S.}\ \bibnamefont
  {Rudner}},\ }\href {http://stacks.iop.org/1367-2630/17/i=12/a=125014}
  {\bibfield  {journal} {\bibinfo  {journal} {New Journal of Physics}\ }\textbf
  {\bibinfo {volume} {17}},\ \bibinfo {pages} {125014} (\bibinfo {year}
  {2015})}\BibitemShut {NoStop}%
\bibitem [{\citenamefont {Asboth}\ \emph {et~al.}(2014)\citenamefont {Asboth},
  \citenamefont {Tarasinski},\ and\ \citenamefont
  {Delplace}}]{PhysRevB.90.125143}%
  \BibitemOpen
  \bibfield  {author} {\bibinfo {author} {\bibfnamefont {J.~K.}\ \bibnamefont
  {Asboth}}, \bibinfo {author} {\bibfnamefont {B.}~\bibnamefont {Tarasinski}},
  \ and\ \bibinfo {author} {\bibfnamefont {P.}~\bibnamefont {Delplace}},\
  }\href@noop {} {\bibfield  {journal} {\bibinfo  {journal} {Phys. Rev. B}\
  }\textbf {\bibinfo {volume} {90}},\ \bibinfo {pages} {125143} (\bibinfo
  {year} {2014})}\BibitemShut {NoStop}%
\bibitem [{\citenamefont {Roy}\ and\ \citenamefont
  {Harper}(2016)}]{Roy2016Floquet}%
  \BibitemOpen
  \bibfield  {author} {\bibinfo {author} {\bibfnamefont {R.}~\bibnamefont
  {Roy}}\ and\ \bibinfo {author} {\bibfnamefont {F.}~\bibnamefont {Harper}},\
  }\href@noop {} {\bibfield  {journal} {\bibinfo  {journal} {arXiv:1603.06944}\
  } (\bibinfo {year} {2016})}\BibitemShut {NoStop}%
\bibitem [{\citenamefont {Shirley}(1965)}]{PhysRev.138.B979}%
  \BibitemOpen
  \bibfield  {author} {\bibinfo {author} {\bibfnamefont {J.~H.}\ \bibnamefont
  {Shirley}},\ }\href {\doibase 10.1103/PhysRev.138.B979} {\bibfield  {journal}
  {\bibinfo  {journal} {Phys. Rev.}\ }\textbf {\bibinfo {volume} {138}},\
  \bibinfo {pages} {B979} (\bibinfo {year} {1965})}\BibitemShut {NoStop}%
\bibitem [{\citenamefont {Sambe}(1973)}]{PhysRevA.7.2203}%
  \BibitemOpen
  \bibfield  {author} {\bibinfo {author} {\bibfnamefont {H.}~\bibnamefont
  {Sambe}},\ }\href {\doibase 10.1103/PhysRevA.7.2203} {\bibfield  {journal}
  {\bibinfo  {journal} {Phys. Rev. A}\ }\textbf {\bibinfo {volume} {7}},\
  \bibinfo {pages} {2203} (\bibinfo {year} {1973})}\BibitemShut {NoStop}%
\bibitem [{\citenamefont {Yan}\ and\ \citenamefont
  {Wang}(2016)}]{PhysRevLett.117.087402}%
  \BibitemOpen
  \bibfield  {author} {\bibinfo {author} {\bibfnamefont {Z.}~\bibnamefont
  {Yan}}\ and\ \bibinfo {author} {\bibfnamefont {Z.}~\bibnamefont {Wang}},\
  }\href {\doibase 10.1103/PhysRevLett.117.087402} {\bibfield  {journal}
  {\bibinfo  {journal} {Phys. Rev. Lett.}\ }\textbf {\bibinfo {volume} {117}},\
  \bibinfo {pages} {087402} (\bibinfo {year} {2016})}\BibitemShut {NoStop}%
\bibitem [{\citenamefont {Wang}\ \emph
  {et~al.}(2016{\natexlab{a}})\citenamefont {Wang}, \citenamefont {Zhou},\ and\
  \citenamefont {Chong}}]{PhysRevB.93.144114}%
  \BibitemOpen
  \bibfield  {author} {\bibinfo {author} {\bibfnamefont {H.}~\bibnamefont
  {Wang}}, \bibinfo {author} {\bibfnamefont {L.}~\bibnamefont {Zhou}}, \ and\
  \bibinfo {author} {\bibfnamefont {Y.~D.}\ \bibnamefont {Chong}},\ }\href
  {\doibase 10.1103/PhysRevB.93.144114} {\bibfield  {journal} {\bibinfo
  {journal} {Phys. Rev. B}\ }\textbf {\bibinfo {volume} {93}},\ \bibinfo
  {pages} {144114} (\bibinfo {year} {2016}{\natexlab{a}})}\BibitemShut
  {NoStop}%
\bibitem [{\citenamefont {Zhou}\ \emph
  {et~al.}(2016{\natexlab{a}})\citenamefont {Zhou}, \citenamefont {Chen},\ and\
  \citenamefont {Gong}}]{PhysRevB.94.075443}%
  \BibitemOpen
  \bibfield  {author} {\bibinfo {author} {\bibfnamefont {L.}~\bibnamefont
  {Zhou}}, \bibinfo {author} {\bibfnamefont {C.}~\bibnamefont {Chen}}, \ and\
  \bibinfo {author} {\bibfnamefont {J.}~\bibnamefont {Gong}},\ }\href {\doibase
  10.1103/PhysRevB.94.075443} {\bibfield  {journal} {\bibinfo  {journal} {Phys.
  Rev. B}\ }\textbf {\bibinfo {volume} {94}},\ \bibinfo {pages} {075443}
  (\bibinfo {year} {2016}{\natexlab{a}})}\BibitemShut {NoStop}%
\bibitem [{\citenamefont {Li}\ \emph {et~al.}(2013)\citenamefont {Li},
  \citenamefont {Wang},\ and\ \citenamefont {Wu}}]{1367-2630-15-8-085002}%
  \BibitemOpen
  \bibfield  {author} {\bibinfo {author} {\bibfnamefont {Y.}~\bibnamefont
  {Li}}, \bibinfo {author} {\bibfnamefont {D.}~\bibnamefont {Wang}}, \ and\
  \bibinfo {author} {\bibfnamefont {C.}~\bibnamefont {Wu}},\ }\href
  {http://stacks.iop.org/1367-2630/15/i=8/a=085002} {\bibfield  {journal}
  {\bibinfo  {journal} {New Journal of Physics}\ }\textbf {\bibinfo {volume}
  {15}},\ \bibinfo {pages} {085002} (\bibinfo {year} {2013})}\BibitemShut
  {NoStop}%
\bibitem [{\citenamefont {Wang}\ \emph
  {et~al.}(2014{\natexlab{b}})\citenamefont {Wang}, \citenamefont {Huang},\
  and\ \citenamefont {Wu}}]{PhysRevB.89.174510}%
  \BibitemOpen
  \bibfield  {author} {\bibinfo {author} {\bibfnamefont {D.}~\bibnamefont
  {Wang}}, \bibinfo {author} {\bibfnamefont {Z.}~\bibnamefont {Huang}}, \ and\
  \bibinfo {author} {\bibfnamefont {C.}~\bibnamefont {Wu}},\ }\href {\doibase
  10.1103/PhysRevB.89.174510} {\bibfield  {journal} {\bibinfo  {journal} {Phys.
  Rev. B}\ }\textbf {\bibinfo {volume} {89}},\ \bibinfo {pages} {174510}
  (\bibinfo {year} {2014}{\natexlab{b}})}\BibitemShut {NoStop}%
\bibitem [{\citenamefont {Ganeshan}\ \emph {et~al.}(2013)\citenamefont
  {Ganeshan}, \citenamefont {Sun},\ and\ \citenamefont
  {Das~Sarma}}]{PhysRevLett.110.180403}%
  \BibitemOpen
  \bibfield  {author} {\bibinfo {author} {\bibfnamefont {S.}~\bibnamefont
  {Ganeshan}}, \bibinfo {author} {\bibfnamefont {K.}~\bibnamefont {Sun}}, \
  and\ \bibinfo {author} {\bibfnamefont {S.}~\bibnamefont {Das~Sarma}},\ }\href
  {\doibase 10.1103/PhysRevLett.110.180403} {\bibfield  {journal} {\bibinfo
  {journal} {Phys. Rev. Lett.}\ }\textbf {\bibinfo {volume} {110}},\ \bibinfo
  {pages} {180403} (\bibinfo {year} {2013})}\BibitemShut {NoStop}%
\bibitem [{\citenamefont {Su}\ \emph {et~al.}(1979)\citenamefont {Su},
  \citenamefont {Schrieffer},\ and\ \citenamefont
  {Heeger}}]{PhysRevLett.42.1698}%
  \BibitemOpen
  \bibfield  {author} {\bibinfo {author} {\bibfnamefont {W.~P.}\ \bibnamefont
  {Su}}, \bibinfo {author} {\bibfnamefont {J.~R.}\ \bibnamefont {Schrieffer}},
  \ and\ \bibinfo {author} {\bibfnamefont {A.~J.}\ \bibnamefont {Heeger}},\
  }\href {\doibase 10.1103/PhysRevLett.42.1698} {\bibfield  {journal} {\bibinfo
   {journal} {Phys. Rev. Lett.}\ }\textbf {\bibinfo {volume} {42}},\ \bibinfo
  {pages} {1698} (\bibinfo {year} {1979})}\BibitemShut {NoStop}%
\bibitem [{\citenamefont {Wakatsuki}\ \emph {et~al.}(2014)\citenamefont
  {Wakatsuki}, \citenamefont {Ezawa}, \citenamefont {Tanaka},\ and\
  \citenamefont {Nagaosa}}]{PhysRevB.90.014505}%
  \BibitemOpen
  \bibfield  {author} {\bibinfo {author} {\bibfnamefont {R.}~\bibnamefont
  {Wakatsuki}}, \bibinfo {author} {\bibfnamefont {M.}~\bibnamefont {Ezawa}},
  \bibinfo {author} {\bibfnamefont {Y.}~\bibnamefont {Tanaka}}, \ and\ \bibinfo
  {author} {\bibfnamefont {N.}~\bibnamefont {Nagaosa}},\ }\href {\doibase
  10.1103/PhysRevB.90.014505} {\bibfield  {journal} {\bibinfo  {journal} {Phys.
  Rev. B}\ }\textbf {\bibinfo {volume} {90}},\ \bibinfo {pages} {014505}
  (\bibinfo {year} {2014})}\BibitemShut {NoStop}%
\bibitem [{\citenamefont {Zeng}\ \emph {et~al.}(2016)\citenamefont {Zeng},
  \citenamefont {Chen},\ and\ \citenamefont {L\"u}}]{Zeng2016}%
  \BibitemOpen
  \bibfield  {author} {\bibinfo {author} {\bibfnamefont {Q.-B.}\ \bibnamefont
  {Zeng}}, \bibinfo {author} {\bibfnamefont {S.}~\bibnamefont {Chen}}, \ and\
  \bibinfo {author} {\bibfnamefont {R.}~\bibnamefont {L\"u}},\ }\href {\doibase
  10.1103/PhysRevB.94.125408} {\bibfield  {journal} {\bibinfo  {journal} {Phys.
  Rev. B}\ }\textbf {\bibinfo {volume} {94}},\ \bibinfo {pages} {125408}
  (\bibinfo {year} {2016})}\BibitemShut {NoStop}%
\bibitem [{\citenamefont {Kitaev}(2001)}]{Kitaev}%
  \BibitemOpen
  \bibfield  {author} {\bibinfo {author} {\bibfnamefont {A.~Y.}\ \bibnamefont
  {Kitaev}},\ }\href@noop {} {\bibfield  {journal} {\bibinfo  {journal}
  {Physics-Uspekhi}\ }\textbf {\bibinfo {volume} {44}},\ \bibinfo {pages} {131}
  (\bibinfo {year} {2001})}\BibitemShut {NoStop}%
\bibitem [{\citenamefont {Tewari}\ and\ \citenamefont
  {Sau}(2012)}]{PhysRevLett.109.150408}%
  \BibitemOpen
  \bibfield  {author} {\bibinfo {author} {\bibfnamefont {S.}~\bibnamefont
  {Tewari}}\ and\ \bibinfo {author} {\bibfnamefont {J.~D.}\ \bibnamefont
  {Sau}},\ }\href@noop {} {\bibfield  {journal} {\bibinfo  {journal} {Phys.
  Rev. Lett.}\ }\textbf {\bibinfo {volume} {109}},\ \bibinfo {pages} {150408}
  (\bibinfo {year} {2012})}\BibitemShut {NoStop}%
\bibitem [{\citenamefont {Wang}\ \emph
  {et~al.}(2016{\natexlab{b}})\citenamefont {Wang}, \citenamefont {Shao},
  \citenamefont {Pan}, \citenamefont {Shen}, \citenamefont {Sheng},\ and\
  \citenamefont {Xing}}]{Wang2016Flux}%
  \BibitemOpen
  \bibfield  {author} {\bibinfo {author} {\bibfnamefont {H.~Q.}\ \bibnamefont
  {Wang}}, \bibinfo {author} {\bibfnamefont {L.~B.}\ \bibnamefont {Shao}},
  \bibinfo {author} {\bibfnamefont {Y.~M.}\ \bibnamefont {Pan}}, \bibinfo
  {author} {\bibfnamefont {R.}~\bibnamefont {Shen}}, \bibinfo {author}
  {\bibfnamefont {L.}~\bibnamefont {Sheng}}, \ and\ \bibinfo {author}
  {\bibfnamefont {D.~Y.}\ \bibnamefont {Xing}},\ }\href@noop {} {\bibfield
  {journal} {\bibinfo  {journal} {Physics Letters A}\ }\textbf {\bibinfo
  {volume} {380}},\ \bibinfo {pages} {3936} (\bibinfo {year}
  {2016}{\natexlab{b}})}\BibitemShut {NoStop}%
\bibitem [{\citenamefont {Heikkil{\"a}}\ \emph {et~al.}(2011)\citenamefont
  {Heikkil{\"a}}, \citenamefont {Kopnin},\ and\ \citenamefont
  {Volovik}}]{Heikkil2011}%
  \BibitemOpen
  \bibfield  {author} {\bibinfo {author} {\bibfnamefont {T.~T.}\ \bibnamefont
  {Heikkil{\"a}}}, \bibinfo {author} {\bibfnamefont {N.~B.}\ \bibnamefont
  {Kopnin}}, \ and\ \bibinfo {author} {\bibfnamefont {G.~E.}\ \bibnamefont
  {Volovik}},\ }\href {\doibase 10.1134/S0021364011150045} {\bibfield
  {journal} {\bibinfo  {journal} {JETP Letters}\ }\textbf {\bibinfo {volume}
  {94}},\ \bibinfo {pages} {233} (\bibinfo {year} {2011})}\BibitemShut
  {NoStop}%
\bibitem [{\citenamefont {Ganeshan}\ and\ \citenamefont
  {Das~Sarma}(2015)}]{PhysRevB.91.125438}%
  \BibitemOpen
  \bibfield  {author} {\bibinfo {author} {\bibfnamefont {S.}~\bibnamefont
  {Ganeshan}}\ and\ \bibinfo {author} {\bibfnamefont {S.}~\bibnamefont
  {Das~Sarma}},\ }\href {\doibase 10.1103/PhysRevB.91.125438} {\bibfield
  {journal} {\bibinfo  {journal} {Phys. Rev. B}\ }\textbf {\bibinfo {volume}
  {91}},\ \bibinfo {pages} {125438} (\bibinfo {year} {2015})}\BibitemShut
  {NoStop}%
\bibitem [{Note1()}]{Note1}%
  \BibitemOpen
  \bibinfo {note} {The sign $\times $ here and in the $Z_{2}\times Z_{2}$ index
  represents a Cartesian product. The elements of $Z\times Z$ or $Z_{2}\times
  Z_{2}$ thus take the form of an ordered pair ($Z1$,$Z2$), where $Z1$ and $Z2$
  are independent of each other.}\BibitemShut {Stop}%
\bibitem [{\citenamefont {Regal}\ \emph {et~al.}(2003)\citenamefont {Regal},
  \citenamefont {Ticknor}, \citenamefont {Bohn},\ and\ \citenamefont
  {Jin}}]{PhysRevLett.90.053201}%
  \BibitemOpen
  \bibfield  {author} {\bibinfo {author} {\bibfnamefont {C.~A.}\ \bibnamefont
  {Regal}}, \bibinfo {author} {\bibfnamefont {C.}~\bibnamefont {Ticknor}},
  \bibinfo {author} {\bibfnamefont {J.~L.}\ \bibnamefont {Bohn}}, \ and\
  \bibinfo {author} {\bibfnamefont {D.~S.}\ \bibnamefont {Jin}},\ }\href
  {\doibase 10.1103/PhysRevLett.90.053201} {\bibfield  {journal} {\bibinfo
  {journal} {Phys. Rev. Lett.}\ }\textbf {\bibinfo {volume} {90}},\ \bibinfo
  {pages} {053201} (\bibinfo {year} {2003})}\BibitemShut {NoStop}%
\bibitem [{\citenamefont {Zhang}\ \emph {et~al.}(2008)\citenamefont {Zhang},
  \citenamefont {Tewari}, \citenamefont {Lutchyn},\ and\ \citenamefont
  {Das~Sarma}}]{PhysRevLett.101.160401}%
  \BibitemOpen
  \bibfield  {author} {\bibinfo {author} {\bibfnamefont {C.}~\bibnamefont
  {Zhang}}, \bibinfo {author} {\bibfnamefont {S.}~\bibnamefont {Tewari}},
  \bibinfo {author} {\bibfnamefont {R.~M.}\ \bibnamefont {Lutchyn}}, \ and\
  \bibinfo {author} {\bibfnamefont {S.}~\bibnamefont {Das~Sarma}},\ }\href
  {\doibase 10.1103/PhysRevLett.101.160401} {\bibfield  {journal} {\bibinfo
  {journal} {Phys. Rev. Lett.}\ }\textbf {\bibinfo {volume} {101}},\ \bibinfo
  {pages} {160401} (\bibinfo {year} {2008})}\BibitemShut {NoStop}%
\bibitem [{\citenamefont {B\"uhler}\ \emph {et~al.}(2014)\citenamefont
  {B\"uhler}, \citenamefont {Lang}, \citenamefont {Kraus}, \citenamefont
  {M\"oller}, \citenamefont {Huber},\ and\ \citenamefont
  {B\"uchler}}]{B2014Majorana}%
  \BibitemOpen
  \bibfield  {author} {\bibinfo {author} {\bibfnamefont {A.}~\bibnamefont
  {B\"uhler}}, \bibinfo {author} {\bibfnamefont {N.}~\bibnamefont {Lang}},
  \bibinfo {author} {\bibfnamefont {C.~V.}\ \bibnamefont {Kraus}}, \bibinfo
  {author} {\bibfnamefont {G.}~\bibnamefont {M\"oller}}, \bibinfo {author}
  {\bibfnamefont {S.~D.}\ \bibnamefont {Huber}}, \ and\ \bibinfo {author}
  {\bibfnamefont {H.~P.}\ \bibnamefont {B\"uchler}},\ }\href@noop {} {\bibfield
   {journal} {\bibinfo  {journal} {Nat. Commun.}\ }\textbf {\bibinfo {volume}
  {5}},\ \bibinfo {pages} {4504} (\bibinfo {year} {2014})}\BibitemShut
  {NoStop}%
\bibitem [{\citenamefont {Verbin}\ \emph {et~al.}(2013)\citenamefont {Verbin},
  \citenamefont {Zilberberg}, \citenamefont {Kraus}, \citenamefont {Lahini},\
  and\ \citenamefont {Silberberg}}]{PhysRevLett.110.076403}%
  \BibitemOpen
  \bibfield  {author} {\bibinfo {author} {\bibfnamefont {M.}~\bibnamefont
  {Verbin}}, \bibinfo {author} {\bibfnamefont {O.}~\bibnamefont {Zilberberg}},
  \bibinfo {author} {\bibfnamefont {Y.~E.}\ \bibnamefont {Kraus}}, \bibinfo
  {author} {\bibfnamefont {Y.}~\bibnamefont {Lahini}}, \ and\ \bibinfo {author}
  {\bibfnamefont {Y.}~\bibnamefont {Silberberg}},\ }\href {\doibase
  10.1103/PhysRevLett.110.076403} {\bibfield  {journal} {\bibinfo  {journal}
  {Phys. Rev. Lett.}\ }\textbf {\bibinfo {volume} {110}},\ \bibinfo {pages}
  {076403} (\bibinfo {year} {2013})}\BibitemShut {NoStop}%
\bibitem [{\citenamefont {Celi}\ \emph {et~al.}(2014)\citenamefont {Celi},
  \citenamefont {Massignan}, \citenamefont {Ruseckas}, \citenamefont {Goldman},
  \citenamefont {Spielman}, \citenamefont {Juzeli\ifmmode~\bar{u}\else
  \={u}\fi{}nas},\ and\ \citenamefont {Lewenstein}}]{PhysRevLett.112.043001}%
  \BibitemOpen
  \bibfield  {author} {\bibinfo {author} {\bibfnamefont {A.}~\bibnamefont
  {Celi}}, \bibinfo {author} {\bibfnamefont {P.}~\bibnamefont {Massignan}},
  \bibinfo {author} {\bibfnamefont {J.}~\bibnamefont {Ruseckas}}, \bibinfo
  {author} {\bibfnamefont {N.}~\bibnamefont {Goldman}}, \bibinfo {author}
  {\bibfnamefont {I.~B.}\ \bibnamefont {Spielman}}, \bibinfo {author}
  {\bibfnamefont {G.}~\bibnamefont {Juzeli\ifmmode~\bar{u}\else
  \={u}\fi{}nas}}, \ and\ \bibinfo {author} {\bibfnamefont {M.}~\bibnamefont
  {Lewenstein}},\ }\href {\doibase 10.1103/PhysRevLett.112.043001} {\bibfield
  {journal} {\bibinfo  {journal} {Phys. Rev. Lett.}\ }\textbf {\bibinfo
  {volume} {112}},\ \bibinfo {pages} {043001} (\bibinfo {year}
  {2014})}\BibitemShut {NoStop}%
\bibitem [{\citenamefont {Chen}\ \emph {et~al.}(2017)\citenamefont {Chen},
  \citenamefont {Mei}, \citenamefont {Su}, \citenamefont {Wang}, \citenamefont
  {Zhu}, \citenamefont {Sheng},\ and\ \citenamefont
  {Xing}}]{0953-8984-29-3-035601}%
  \BibitemOpen
  \bibfield  {author} {\bibinfo {author} {\bibfnamefont {M.~N.}\ \bibnamefont
  {Chen}}, \bibinfo {author} {\bibfnamefont {F.}~\bibnamefont {Mei}}, \bibinfo
  {author} {\bibfnamefont {W.}~\bibnamefont {Su}}, \bibinfo {author}
  {\bibfnamefont {H.-Q.}\ \bibnamefont {Wang}}, \bibinfo {author}
  {\bibfnamefont {S.-L.}\ \bibnamefont {Zhu}}, \bibinfo {author} {\bibfnamefont
  {L.}~\bibnamefont {Sheng}}, \ and\ \bibinfo {author} {\bibfnamefont {D.~Y.}\
  \bibnamefont {Xing}},\ }\href
  {http://stacks.iop.org/0953-8984/29/i=3/a=035601} {\bibfield  {journal}
  {\bibinfo  {journal} {Journal of Physics: Condensed Matter}\ }\textbf
  {\bibinfo {volume} {29}},\ \bibinfo {pages} {035601} (\bibinfo {year}
  {2017})}\BibitemShut {NoStop}%
\bibitem [{\citenamefont {Niu}\ \emph {et~al.}(2012)\citenamefont {Niu},
  \citenamefont {Chung}, \citenamefont {Hsu}, \citenamefont {Mandal},
  \citenamefont {Raghu},\ and\ \citenamefont
  {Chakravarty}}]{PhysRevB.85.035110}%
  \BibitemOpen
  \bibfield  {author} {\bibinfo {author} {\bibfnamefont {Y.}~\bibnamefont
  {Niu}}, \bibinfo {author} {\bibfnamefont {S.~B.}\ \bibnamefont {Chung}},
  \bibinfo {author} {\bibfnamefont {C.-H.}\ \bibnamefont {Hsu}}, \bibinfo
  {author} {\bibfnamefont {I.}~\bibnamefont {Mandal}}, \bibinfo {author}
  {\bibfnamefont {S.}~\bibnamefont {Raghu}}, \ and\ \bibinfo {author}
  {\bibfnamefont {S.}~\bibnamefont {Chakravarty}},\ }\href {\doibase
  10.1103/PhysRevB.85.035110} {\bibfield  {journal} {\bibinfo  {journal} {Phys.
  Rev. B}\ }\textbf {\bibinfo {volume} {85}},\ \bibinfo {pages} {035110}
  (\bibinfo {year} {2012})}\BibitemShut {NoStop}%
\bibitem [{\citenamefont {Meng}\ and\ \citenamefont
  {Balents}(2012)}]{PhysRevB.86.054504}%
  \BibitemOpen
  \bibfield  {author} {\bibinfo {author} {\bibfnamefont {T.}~\bibnamefont
  {Meng}}\ and\ \bibinfo {author} {\bibfnamefont {L.}~\bibnamefont {Balents}},\
  }\href {\doibase 10.1103/PhysRevB.86.054504} {\bibfield  {journal} {\bibinfo
  {journal} {Phys. Rev. B}\ }\textbf {\bibinfo {volume} {86}},\ \bibinfo
  {pages} {054504} (\bibinfo {year} {2012})}\BibitemShut {NoStop}%
\bibitem [{\citenamefont {Cho}\ \emph {et~al.}(2012)\citenamefont {Cho},
  \citenamefont {Bardarson}, \citenamefont {Lu},\ and\ \citenamefont
  {Moore}}]{PhysRevB.86.214514}%
  \BibitemOpen
  \bibfield  {author} {\bibinfo {author} {\bibfnamefont {G.~Y.}\ \bibnamefont
  {Cho}}, \bibinfo {author} {\bibfnamefont {J.~H.}\ \bibnamefont {Bardarson}},
  \bibinfo {author} {\bibfnamefont {Y.-M.}\ \bibnamefont {Lu}}, \ and\ \bibinfo
  {author} {\bibfnamefont {J.~E.}\ \bibnamefont {Moore}},\ }\href {\doibase
  10.1103/PhysRevB.86.214514} {\bibfield  {journal} {\bibinfo  {journal} {Phys.
  Rev. B}\ }\textbf {\bibinfo {volume} {86}},\ \bibinfo {pages} {214514}
  (\bibinfo {year} {2012})}\BibitemShut {NoStop}%
\bibitem [{\citenamefont {Bednik}\ \emph {et~al.}(2015)\citenamefont {Bednik},
  \citenamefont {Zyuzin},\ and\ \citenamefont {Burkov}}]{PhysRevB.92.035153}%
  \BibitemOpen
  \bibfield  {author} {\bibinfo {author} {\bibfnamefont {G.}~\bibnamefont
  {Bednik}}, \bibinfo {author} {\bibfnamefont {A.~A.}\ \bibnamefont {Zyuzin}},
  \ and\ \bibinfo {author} {\bibfnamefont {A.~A.}\ \bibnamefont {Burkov}},\
  }\href {\doibase 10.1103/PhysRevB.92.035153} {\bibfield  {journal} {\bibinfo
  {journal} {Phys. Rev. B}\ }\textbf {\bibinfo {volume} {92}},\ \bibinfo
  {pages} {035153} (\bibinfo {year} {2015})}\BibitemShut {NoStop}%
\bibitem [{\citenamefont {Zhou}\ \emph
  {et~al.}(2016{\natexlab{b}})\citenamefont {Zhou}, \citenamefont {Gao},\ and\
  \citenamefont {Wang}}]{PhysRevB.93.094517}%
  \BibitemOpen
  \bibfield  {author} {\bibinfo {author} {\bibfnamefont {T.}~\bibnamefont
  {Zhou}}, \bibinfo {author} {\bibfnamefont {Y.}~\bibnamefont {Gao}}, \ and\
  \bibinfo {author} {\bibfnamefont {Z.~D.}\ \bibnamefont {Wang}},\ }\href
  {\doibase 10.1103/PhysRevB.93.094517} {\bibfield  {journal} {\bibinfo
  {journal} {Phys. Rev. B}\ }\textbf {\bibinfo {volume} {93}},\ \bibinfo
  {pages} {094517} (\bibinfo {year} {2016}{\natexlab{b}})}\BibitemShut
  {NoStop}%
\bibitem [{\citenamefont {Wang}\ \emph
  {et~al.}(2016{\natexlab{c}})\citenamefont {Wang}, \citenamefont {Hao},
  \citenamefont {Wang},\ and\ \citenamefont {Ting}}]{PhysRevB.93.184511}%
  \BibitemOpen
  \bibfield  {author} {\bibinfo {author} {\bibfnamefont {R.}~\bibnamefont
  {Wang}}, \bibinfo {author} {\bibfnamefont {L.}~\bibnamefont {Hao}}, \bibinfo
  {author} {\bibfnamefont {B.}~\bibnamefont {Wang}}, \ and\ \bibinfo {author}
  {\bibfnamefont {C.~S.}\ \bibnamefont {Ting}},\ }\href {\doibase
  10.1103/PhysRevB.93.184511} {\bibfield  {journal} {\bibinfo  {journal} {Phys.
  Rev. B}\ }\textbf {\bibinfo {volume} {93}},\ \bibinfo {pages} {184511}
  (\bibinfo {year} {2016}{\natexlab{c}})}\BibitemShut {NoStop}%
\bibitem [{\citenamefont {Zhou}\ \emph {et~al.}(2014)\citenamefont {Zhou},
  \citenamefont {Wang}, \citenamefont {Ho},\ and\ \citenamefont
  {Gong}}]{Zhou2014}%
  \BibitemOpen
  \bibfield  {author} {\bibinfo {author} {\bibfnamefont {L.}~\bibnamefont
  {Zhou}}, \bibinfo {author} {\bibfnamefont {H.}~\bibnamefont {Wang}}, \bibinfo
  {author} {\bibfnamefont {Y.~D.}\ \bibnamefont {Ho}}, \ and\ \bibinfo {author}
  {\bibfnamefont {J.}~\bibnamefont {Gong}},\ }\href {\doibase
  10.1140/epjb/e2014-50465-9} {\bibfield  {journal} {\bibinfo  {journal} {The
  European Physical Journal B}\ }\textbf {\bibinfo {volume} {87}},\ \bibinfo
  {pages} {1} (\bibinfo {year} {2014})}\BibitemShut {NoStop}%
\bibitem [{\citenamefont {Thouless}(1983)}]{PhysRevB.27.6083}%
  \BibitemOpen
  \bibfield  {author} {\bibinfo {author} {\bibfnamefont {D.~J.}\ \bibnamefont
  {Thouless}},\ }\href {\doibase 10.1103/PhysRevB.27.6083} {\bibfield
  {journal} {\bibinfo  {journal} {Phys. Rev. B}\ }\textbf {\bibinfo {volume}
  {27}},\ \bibinfo {pages} {6083} (\bibinfo {year} {1983})}\BibitemShut
  {NoStop}%
\bibitem [{\citenamefont {Ho}\ and\ \citenamefont
  {Gong}(2012)}]{PhysRevLett.109.010601}%
  \BibitemOpen
  \bibfield  {author} {\bibinfo {author} {\bibfnamefont {D.~Y.~H.}\
  \bibnamefont {Ho}}\ and\ \bibinfo {author} {\bibfnamefont {J.}~\bibnamefont
  {Gong}},\ }\href {\doibase 10.1103/PhysRevLett.109.010601} {\bibfield
  {journal} {\bibinfo  {journal} {Phys. Rev. Lett.}\ }\textbf {\bibinfo
  {volume} {109}},\ \bibinfo {pages} {010601} (\bibinfo {year}
  {2012})}\BibitemShut {NoStop}%
\bibitem [{\citenamefont {Keselman}\ \emph {et~al.}(2013)\citenamefont
  {Keselman}, \citenamefont {Fu}, \citenamefont {Stern},\ and\ \citenamefont
  {Berg}}]{PhysRevLett.111.116402}%
  \BibitemOpen
  \bibfield  {author} {\bibinfo {author} {\bibfnamefont {A.}~\bibnamefont
  {Keselman}}, \bibinfo {author} {\bibfnamefont {L.}~\bibnamefont {Fu}},
  \bibinfo {author} {\bibfnamefont {A.}~\bibnamefont {Stern}}, \ and\ \bibinfo
  {author} {\bibfnamefont {E.}~\bibnamefont {Berg}},\ }\href {\doibase
  10.1103/PhysRevLett.111.116402} {\bibfield  {journal} {\bibinfo  {journal}
  {Phys. Rev. Lett.}\ }\textbf {\bibinfo {volume} {111}},\ \bibinfo {pages}
  {116402} (\bibinfo {year} {2013})}\BibitemShut {NoStop}%
\end{thebibliography}%

\end{document}